\documentclass[aps,twocolumn,pre,superscriptaddress]{revtex4}
\usepackage{amsmath}
\usepackage{amssymb}
\usepackage{graphicx}
\usepackage{color}

\begin{document}

\title{Ott-Antonsen ansatz is the only admissible truncation of a circular cumulant series}

\author{Denis S.\ Goldobin}
\affiliation{Institute of Continuous Media Mechanics, UB RAS, Acad.\ Korolev Street 1,
 614013\,Perm, Russia}
\affiliation{Department of Theoretical Physics, Perm State University, Bukirev Street 15,
 614990\,Perm, Russia}
\author{Anastasiya V.\ Dolmatova}
\affiliation{Institute of Continuous Media Mechanics, UB RAS, Acad.\ Korolev Street 1,
 614013\,Perm, Russia}
\affiliation{Institute for Information Transmission Problems, RAS, B.\,Karetny\,per.\,19,
 127051\,Moscow, Russia}
\date{\today}

\begin{abstract}
The cumulant representation is common in classical statistical physics for variables on the real line and the issue of closures of cumulant expansions is well elaborated. The case of phase variables significantly differs from the case of linear ones; the relevant order parameters are the Kuramoto-Daido ones but not the conventional moments. One can formally introduce `circular' cumulants for Kuramoto-Daido order parameters, similar to the conventional cumulants for moments. The circular cumulant expansions allow to advance beyond the Ott-Antonsen theory and consider populations of real oscillators.
First, we show that truncation of circular cumulant expansions, except for the Ott-Antonsen case, is forbidden.
Second, we compare this situation to the case of the Gaussian distribution of a linear variable, where the second cumulant is nonzero and all the higher cumulants are zero, and elucidate why keeping up to the second cumulant is admissible for a linear variable, but forbidden for circular cumulants.
Third, we discuss the implication of this truncation issue to populations of quadratic integrate-and-fire neurons [E.\ Montbri\'o, D.\ Paz\'o, A.\ Roxin, Phys.\ Rev.\ X {\bf 5}, 021028 (2015)], where within the framework of macroscopic description, the firing rate diverges for any finite truncation of the cumulant series, and discuss how one should handle these situations.
Fourth, we consider the cumulant-based low-dimensional reductions for macroscopic population dynamics in the context of this truncation issue. These reductions are applicable, where the cumulant series exponentially decay with the cumulant order, i.e., they form a geometric progression hierarchy. Fifth, we demonstrate the formation of this hierarchy for generic distributions on the circle and experimental data for coupled biological and electrochemical oscillators.
Our main conclusion for applications is that, if the first and second circular cumulants are nonzero, there must be infinitely many nonzero higher cumulants. However, these higher cumulants can be small, in which case one can construct approximate equations for the dynamics of a finite number of leading cumulants.

\vspace{10pt}
\hspace{7cm}
\begin{tabular}{cl}
Subject Areas: & Statistical Physics,\\
& Interdisciplinary Physics,\\
& Nonlinear Dynamics
\end{tabular}
\end{abstract}

\maketitle

\section{Introduction}
The problem of cumulant representation and closure of cumulant expansions is likely one of the most generic problems in non-equilibrium statistical physics.
For instance, starting with the Boltzmann kinetic equation governing dynamics of the one-particle probability density function, one can find an infinite chain of equations for moments of the probability density with respect to velocity: mass density, momentum density, the second moment, etc.~\cite{Pitaevskii-Lifshitz-1981}. For an isotropic medium, one deals with the convolution of the second moment, which is the net mechanical energy. Equations for these moments are nothing else but the conservation laws for mass, momentum, and net mechanical energy. Moreover, it is conventional to deduct the macroscopic velocity contributions from the second moment and obtain the energy conservation equation in terms of the internal energy; mathematically, this deducting corresponds to switching from moments to cumulants. Closure of the equation chain is a trivial task: for incompressible flows one makes this closure by assuming a constant density and discarding the energy equation (the second cumulant), for compressible ones the third cumulant is neglected and an algebraic relation between pressure and internal energy is adopted. These closures are correct for weakly non-equilibrium systems (mathematically, the limit of small Knudsen number), which is relevant for nearly all macroscopic processes on the Earth's surface with excellent accuracy. Thus, the equations of continuous media mechanics are actually an example of the cumulant expansion and its closure.

The two-cumulant representation corresponding to the case of continuous media mechanics is actually a Gaussian approximation for the microscopic velocity distribution. However, in statistical physics, one can address the problem of macroscopic description for exotic systems, where particles are not actual molecules or atoms, but macroscopic objects: grains, stones, asteroids, etc. For these systems not only may the limit of small Knudsen number not be relevant, but also the reversibility of interparticle collisions is lost, leading to essentially non-Gaussian distributions of the microscopic velocity~\cite{Brilliantov-Poeschel-2010}. In this case, one has to deal with higher-order cumulants and look for non-Gaussian closures.

The problem of adiabatic velocity elimination for Brownian particles belongs to the same class of problems~\cite{Wilemski-1976,Gardiner-1983}. Indeed, the derivation of the enhanced Smoluchowski equation for the probability density can be conducted as constructing an expansion in cumulants with respect to velocity and closure for higher-order cumulants within the limit of small inertia.

Cumulant expansions and closures either with Gaussian approximation or accounting for the higher cumulants are quite abundant in theoretical studies on stochastic systems (e.g., \cite{Wilemski-1976,Pikovsky-etal-2002,Zaks-etal-2003,Sonnenschein-Schimansky-Geier-2013,Sonnenschein-etal-2015}).

Simultaneously, it turns out that one has to be subtle with high-order cumulant approximations. The only physically meaningful case with a finite number of nonzero cumulants is the case of Gaussian distribution, where the first and second cumulants can be nonzero. With any nonzero cumulant of higher order, the series of nonzero elements becomes infinite~\cite{Lukacs-1970}. Furthermore, one can derive the Fokker--Planck equation for white Gaussian noise~\cite{Gardiner-1983}, but an analogous equation for any other finite number of nonzero cumulants will exhibit unphysical dynamics. Nonetheless, it is possible and quite common to be able to benefit from accounting for the impact of third and fourth cumulants on the dynamics of the first and second ones, while the corrections owned by the higher-order cumulants are neglected.

Classical statistical physics deals with variables on the real line. Studies on self-organization in active media and control theory, however, revealed the practical and theoretical importance of phase variables defined on the circle~\cite{Winfree-1967,Kuramoto-2003,Pikovsky-Rosenblum-Kurths-2003}. With phase variable $\varphi$, the conventional moments are poor representatives of the macroscopic order, while the Kuramoto--Daido parameters $Z_m=\langle{e^{im\varphi}}\rangle$ \cite{Daido-1996} become a natural measure of the order.

A significant breakthrough in the theory of collective phenomena in populations of phase elements---which can be limit-cycle oscillators or excitable elements---was based on the Ott--Antonsen (OA) theory~\cite{Ott-Antonsen-2008,Ott-Antonsen-2009} related to an important particular case of the Watanabe--Stro\-gatz theory~\cite{Watanabe-Strogatz-1993,Watanabe-Strogatz-1994,Pikovsky-Rosenblum-2008,Marvel-Mirollo-Strogatz-2009}. The OA theory provided the closure $Z_m=(Z_1)^m$ for infinite equation chains for the Kuramoto--Daido order parameters and allowed one to obtain a self-contained dynamics equation for $Z_1$.
This closure can be referred to as the Ott--Antonsen ansatz; the issue of the attractivity of the manifold $Z_m=(Z_1)^m$ was also elaborated within the theory.
Recently~\cite{Tyulkina-etal-2018}, it was proposed to treat the order parameters $Z_m$ as moments and deal with the formally corresponding `circular cumulants'. The circular cumulant approach allowed to go beyond the OA ansatz~\cite{Tyulkina-etal-2018,Goldobin-etal-2018,Tyulkina-etal-2019,Tyulkina-Goldobin-Pikovsky-2019} and handle cases where the genuine OA theory is inapplicable~\cite{Tyulkina-etal-2018,Goldobin-etal-2018}. Within the framework of the circular cumulant approach, the OA theory turned out to be the case where only the first circular cumulant is nonzero. In~\cite{Tyulkina-etal-2018,Goldobin-etal-2018}, corrections owned by the second cumulant allowed to achieve accurate results where the OA ansatz was significantly inaccurate. The cumulant approach can be also applicable for a theoretical analysis of the non-OA situations, e.g., in \cite{Laing-2012,Laing-2018b,Franovic-Klinshov-2019,Pelka-Peano-Xuereb-2019,Ratas-Pyragas-2019,Goldobin-Dolmatova-2019b}.

The circular cumulant approach can be also employed in statistical physics problems of directional systems~\cite{Ley-Verdebout-2017}: magnetic nanoparticle ensembles, liquid crystals, active Brownian particles, optomechanical systems, etc.

The retrospective on the statistical physics experience with linear variables suggests, that the progress in the theory of collective phenomena and self-organization in oscillatory and excitable media can be closely interwoven with implementation of the circular cumulant representations. The practical application of the circular cumulant approach and closures with a finite number of cumulants raises the questions of (i)~physically admissible truncations of the cumulant series and (ii)~dealing with closures where a finite number of circular cumulants is kept.

In this paper, we derive that, in contrast to the linear variable case, the only physically meaningful truncation is the single-cumulant one, which corresponds to the wrapped Cauchy distribution of phases and lies at the basis of the Ott--Antonsen ansatz. With more than one nonzero circular cumulant, the cumulant series of physically meaningful distributions must be infinite. Nonetheless, similarly to the case of linear variables, where for non-Gaussian cases one can benefit from corrections owned by a finite number of higher-order cumulants, closures with more than one circular cumulant are useful. We also show that in some physical systems the macroscopic variables driving collective dynamics (e.g., neuron firing rate~\cite{Pazo-Montbrio-2014,Montbrio-Pazo-Roxin-2015,Pazo-Montbrio-2016,Devalle-Pazo-Montbrio-2018,Volo-Torcini-2018,Ratas-Pyragas-2019}) can depend on the Kuramoto--Daido order parameters in such a way that a careless representation of these variables in terms of circular cumulants can be always diverging. Further, we discuss how the issue of approximations with a finite number of circular cumulants should be handled in a regular way in the cases where the cumulants form a decaying geometric progression. We report the presence of this progression for important generic distributions and demonstrate it with experimental data for coupled biological and electrochemical oscillators.

\section{Truncated circular cumulant series}
\label{sec:Zm}
\subsection{Ott--Antonsen ansatz as a one-cumulant truncation}
\label{sec:OA}
In this subsection we give a brief introduction to the Ott--Antonsen theory and reformulate it in terms of circular cumulants. Basically, it is formulated for populations of identical phase elements governed by equations
\begin{equation}
\dot\varphi_j=\omega(t)+\mathrm{Im}(2h(t)e^{-i\varphi_j})\,,
%\qquad j=1,...,N,
\label{eq:OA1}
\end{equation}
where $\omega(t)$ and $h(t)$ are arbitrary real- and complex-valued functions of time.
The theory is valid in the thermodynamic limit of an infinitely large population, where the system state is naturally represented by the probability density function of phases $w(\varphi,t)$. The master equation for $w(\varphi,t)$ reads
\begin{equation}
\frac{\partial w}{\partial t} +\frac{\partial}{\partial\varphi}\Big(\left(\omega(t) -ih(t)e^{-i\varphi}+ih^\ast(t)e^{i\varphi}\right)w\Big)=0\,.
\label{eq:OA2}
\end{equation}
In Fourier space, where
\begin{equation}
w(\varphi,t)=\frac{1}{2\pi}\Big[1+\sum_{m=1}^\infty \left(Z_m(t)e^{-im\varphi}+Z_m^\ast(t)e^{im\varphi}\right)\Big],
\label{eq:OA3}
\end{equation}
master equation~(\ref{eq:OA2}) takes the form
\begin{equation}
\dot{Z}_m=im\omega Z_m+mhZ_{m-1}-mh^\ast{Z}_{m+1}\,,
\label{eq:OA4}
\end{equation}
where $Z_0=1$ and $Z_{-m}=Z_m^\ast$ by definition. Ott and Antonsen~\cite{Ott-Antonsen-2008} noticed that Eq.~(\ref{eq:OA4}) admits the solution $Z_m(t)=\left[Z_1(t)\right]^m$ with the order parameter $Z_1=\langle{e^{i\varphi}}\rangle$ obeying a simple self-contained equation:
\begin{equation}
\dot{Z}_1=i\omega Z_1+h-h^\ast{Z}_1^2\,.
\label{eq:OA5}
\end{equation}

The OA manifold $Z_m=\left(Z_1\right)^m$ is neutrally stable for perfectly identical population elements, but becomes weakly attracting for typical cases of imperfect parameter identity, where the parameter distribution is continuous \cite{Ott-Antonsen-2009,Mirollo-2012,Pietras-Daffertshofer-2016}. Eq.~(\ref{eq:OA5}) is an exact result, which provides a closed equation for the dynamics of order parameter $Z_1$ and made a ground for a significant advance in various studies on collective phenomena.

For $Z_m=\left(Z_1\right)^m$ with $Z_1=R\,e^{i\psi}$, $R=|Z_1|$, one can calculate the probability density~(\ref{eq:OA3})
\begin{equation}
w_\mathrm{OA}(\varphi)=\frac{1-|Z_1|^2}{2\pi|1-Z_1e^{-i\varphi}|^2} =\frac{(2\pi)^{-1}(1-R^2)}{1-R\cos(\varphi-\psi)}
\label{eq:OA6}
\end{equation}
and find that it is a wrapped Cauchy distribution
\begin{equation}
w_\mathrm{wC}(\varphi)=\sum_{n=-\infty}^{+\infty}
 \frac{\pi^{-1}\gamma}{(\theta+2\pi n)^2+\gamma^2} =\frac{1}{2\pi}\frac{1-e^{-2\gamma}}{1-e^{-\gamma}\cos\theta}
\nonumber%\label{eq:OA7}
\end{equation}
with $\theta=\varphi-\psi$ and $Z_1=e^{-\gamma+i\psi}$.

Let us now consider $Z_m$ as moments of $e^{i\varphi}$ and formally introduce corresponding cumulants~\cite{Tyulkina-etal-2018}. The latter quantities are not conventional cumulants of original variable $\varphi$; hence, we are free to choose the normalization for them and will refer to them as `circular cumulants'. With the moment generating function
\begin{equation}
F(k)\equiv\langle\exp(ke^{i\varphi})\rangle=1+Z_1k+Z_2\frac{k^2}{2!}+Z_3\frac{k^3}{3!}+\dots
\label{eq:OA7}
\end{equation}
we define circular cumulants $\kappa_m$ via generating function
\begin{equation}
\Psi(k)\equiv k\frac{\partial}{\partial k}\ln{F(k)}\equiv\kappa_1k+\kappa_2k^2+\kappa_3k^3+\dots\,.
\label{eq:OA8}
\end{equation}
For instance, the first three circular cumulants are
\[
\kappa_1=Z_1,\quad
\kappa_2=Z_2-Z_1^2,\quad
\kappa_3=\frac{Z_3-3Z_2Z_1+2Z_1^3}{2}.
\]

In terms of circular cumulants, the OA manifold $Z_m=(Z_1)^m$ acquires a simple form:
\begin{equation}
\kappa_1=Z_1\,,\qquad \kappa_{m\ge2}=0\,.
\label{eq:OA9}
\end{equation}
Thus, the Ott--Antonsen ansatz {\it or} the case of a wrapped Cauchy distribution of phases can be considered as the {\em one-cumulant truncation} of a circular cumulant series.

Equation system~(\ref{eq:OA4}) turns into
\begin{align}
\dot{\kappa}_m&=im\omega\kappa_m+h\delta_{1m}
\nonumber\\
&\qquad
-mh^\ast\Big(m\kappa_{m+1}+\sum_{j=1}^m\kappa_{m-j+1}\kappa_j\Big),
\label{eq:OA10}
\end{align}
where $\delta_{1m}=1$ for $m=1$ and $0$ otherwise~\cite{Tyulkina-etal-2018}. One can employ Eqs.~(\ref{eq:OA10}) for studying the population dynamics beyond the OA ansatz and derive analytically solvable extensions of the OA solution~\cite{Tyulkina-etal-2019}. For the systems where the OA form~(\ref{eq:OA1}) of equations is violated, within the framework of the circular cumulant approach, one can derive modified versions of equation system~(\ref{eq:OA10}) and low-dimensional equation systems for order parameters (e.g.,~\cite{Tyulkina-etal-2018,Goldobin-Dolmatova-2019b,Ratas-Pyragas-2019}).

\subsection{Two-cumulant truncation}
\label{ssec:2cZm}
It is instructive to start the analysis with the truncation where only two first circular cumulants are nonzero.
For a detailed step-by-step derivation see Appendix~\ref{asec:2cZm}, while here we pin the principal derivation points and present its results.
% (Appendix is organized in such a way, that one can read it before reading this section.)

%For a characteristic function of a random phase variable $\varphi$, $F(k)\equiv\langle\exp(ke^{i\varphi})\rangle=1+Z_1k+Z_2\frac{k^2}{2!}+Z_3\frac{k^3}{3!}+\dots$, it is convenient to define circular cumulants $\kappa_m$ via the generating function $\Psi(k)=k\partial_k\ln{F(k)}=\kappa_1k+\kappa_2k^2+\kappa_3k^3+\dots$ \cite{Tyulkina-etal-2018}.

With only two first nonzero $\kappa_m$, the circular cumulant generating function $\Psi(k)=\kappa_1k+\kappa_2k^2$, and $\ln{F}=\kappa_1k+\kappa_2\frac{k^2}{2}$. Thus,
\begin{align}
&\textstyle
F=e^{\kappa_1k}e^{\kappa_2\frac{k^2}{2}}
%\nonumber\\
%&\quad\textstyle
%{}
=\sum\limits_{m_1=0}^{\infty}\kappa_1^{m_1}\frac{k^{m_1}}{m_1!} \sum\limits_{m_2=0}^{\infty}\big(\frac{\kappa_2}{2}\big)^{m_2}\frac{k^{2m_2}}{m_2!}\,.
\label{eq:t101}
\end{align}
Gathering the terms with $k^m$ in product~(\ref{eq:t101}), one finds
\begin{equation}
Z_m=\!\sum_{j=0}^{\mathrm{int}(m/2)}\!\frac{m!}{(m-2j)!j!}\frac{\kappa_1^{m-2j}\kappa_2^j}{2^j}
\equiv\!\sum_{j=0}^{\mathrm{int}(m/2)}\!s_{m;j}\,,
\label{eq:t1023}
\end{equation}
where $\mathrm{int}(\cdot)$ returns the integer part of a number and $s_{m;j}$ is a brief notation for the sum terms.

For arbitrary nonzero $\kappa_1$ and $\kappa_2$, one can specify sufficiently large $m$ such that the dominating contribution into the sum will be made by summands $s_{m;j}$ which are far from the sum edges. Moreover, the absolute value of the summands $|s_{m;j}|$ will be modulated by a Gaussian function of index $j$. For large $m$, $j$, and $(m-2j)$, the Stirling's approximation, $n!\approx\sqrt{2\pi n}(n/e)^n$, can be employed for calculation of the summands.

Calculations take the simplest form for
\[
m\gg M_3\equiv a^{-3}+a^3,\qquad a\equiv\frac{|\kappa_1|^2}{2|\kappa_2|}\,.
\]
In this case, one finds the summand with the maximal absolute value at $j=l$:
\begin{equation}
2l=m-\sqrt{2ma}+a+\frac12+o(1)\,.
\label{eq:t106}
\end{equation}
The absolute value
\begin{equation}
|s_{m;l}|=
\frac{1+o(1)}{\sqrt{\pi}(2ma)^\frac14} \left(\frac{m|\kappa_2|}{e}\right)^{\frac{m}{2}}e^{\sqrt{2ma}-\frac{a}{2}}\,,
\label{eq:t108}
\end{equation}
and for the neighboring terms, one can calculate
\begin{equation}
\frac{s_{m;l+r}}{s_{m;l}}\approx e^{i\Theta r-\sqrt{\frac{2}{ma}}r^2},
\label{eq:t109}
\end{equation}
where
\[
e^{i\Theta}\equiv\frac{\kappa_2}{|\kappa_2|}\frac{|\kappa_1|^2}{\kappa_1^2}\,.
\]

As the Gaussian function in Eq.~(\ref{eq:t109}) is localized on the scale $(ma)^{1/4}\gg1$, the summand magnitude slowly varies with the index $r$, and one can assess the order of magnitude of the sum in $Z_m$ as an integral;
\[
\textstyle
Z_m\approx s_{m;l}\int\limits_{-\infty}^{+\infty}\frac{s_{m;l+r}}{s_{m;l}}\mathrm{d}r
=\sqrt{\pi}\left(\frac{ma}{2}\right)^\frac14 e^{-\frac{\sqrt{2ma}\Theta^2}{16}}s_{m;l}\,.
\]
Although the transition from a sum to an integral introduces inaccuracy for finite $\Theta$, the results with an integral match the exact sum~(\ref{eq:t1023}) surprisingly well (see Fig.~\ref{fig6} in Appendix).
Substituting $|s_{m;l}|$ from Eq.~(\ref{eq:t108}), one obtains
\begin{align}
&\textstyle
|Z_m|\approx\frac{1}{\sqrt{2}}\left(\frac{m|\kappa_2|}{e}\right)^\frac{m}{2}
e^{\frac{\sqrt{m}|\kappa_1|}{\sqrt{|\kappa_2|}}\left(1-\frac{\Theta^2}{16}\right) -\frac{|\kappa_1|^2}{4|\kappa_2|}}.
\label{eq:t110}
\end{align}

From Eq.~(\ref{eq:t110}), which is valid for $m\gg M_3$, one can see that for $m>e/|\kappa_2|$, the absolute value $|Z_m|$ becomes larger than $1$. However, this is not possible for the average value $\langle{e^{im\varphi}}\rangle$ of a phase $\varphi$ on the circle. Technically, for the distribution density $w(\varphi)$, the condition $|\int_0^{2\pi}\!w(\varphi)e^{im\varphi}\mathrm{d}\varphi\,|>1$ under the normalization condition $\int_0^{2\pi}\!w(\varphi)\,\mathrm{d}\varphi=1$ requires negativity of $w(\varphi)$ for some $\varphi$.

\subsection{Truncation of $\kappa_{m}$ for $m>N$}
\label{ssec:NcZm}
Similarly to the previous subsection, a detailed step-by-step derivation is provided in Appendix~\ref{asec:NcZm}, while here we pin the principal derivation points and present its results.
For nonzero $\kappa_m$, $m=1,2,...,N$,
\begin{align}
%\textstyle
F(k)=\exp\left(\sum_{m=1}^N\kappa_m\frac{k^m}{m}\right)
=\prod\limits_{m=1}^Ne^\frac{\kappa_mk^m}{m}
\nonumber\\
%\textstyle
=\prod\limits_{m=1}^N\sum_{j_m=0}^\infty\left(\frac{\kappa_m}{m}\right)^{j_m}\frac{k^{m j_m}}{(j_m)!}\,;
\nonumber
\end{align}
therefore,
\begin{equation}
%\textstyle
Z_m=\!\!\!\sum\limits_{j_1+2j_2+\dots \atop {}+Nj_N=m}\!
 \frac{m!}{j_1!j_2!\dots j_N!}\left(\frac{\kappa_1}{1}\right)^{j_1} \left(\frac{\kappa_2}{2}\right)^{j_2}\!\!\dots
 \left(\frac{\kappa_N}{N}\right)^{j_N}\!\!.
\label{eq:t201}
\end{equation}

For sufficiently large $m$, the principal contributions are owned by the summands which are far from the boundaries of the summation domain in the index space. For large $m$, $j_1$, $j_2$,..., $j_N$, one can use the Stirling's formula and replace the summation with integration over the hyperplane $j_1+2j_2+3j_3+\dots+Nj_N=m$;
\begin{align}
&\textstyle
Z_m\approx\int\mathrm{d}j_2\int\mathrm{d}j_3\dots\int\mathrm{d}j_{N} s_{m;j_1j_2\dots j_N}\,,
\label{eq:t202a}
%\\
%&\textstyle
%s_{m;j_1j_2\dots j_N}\equiv
%\sqrt{\frac{m}{(2\pi)^{N-1}j_1j_2\dots j_N}}
%\nonumber\\
%&\textstyle\qquad\qquad\qquad
% \times\frac{m^m\kappa_1^{j_1}\kappa_2^{j_2}\dots\kappa_N^{j_N}}{(j_1)^{j_1}(2ej_2)^{j_2} \dots(Ne^{N-1}j_N)^{j_N}},
%\label{eq:t202b}
\end{align}
where $j_1=m-2j_2-3j_3-\dots-Nj_N$. The summand magnitude will be as well modulated by a Gaussian function of the indices.

One can find the maximum of $|s_{m;j_1j_2\dots j_N}|$ on the hyperplane by means of the method of Lagrange multipliers.
To the leading order, one finds the maximum position
\begin{equation}
\textstyle
nl_n\approx\frac{|\kappa_n|}{|\kappa_N|^\frac{n}{N}}m^\frac{n}{N} -\frac{n}{N}\frac{|\kappa_n||\kappa_{N-1}|}{|\kappa_N|^\frac{N+n-1}{N}}m^\frac{n-1}{N}\,,
\label{eq:t205}
\end{equation}
and
\begin{align}
&|s_{m;l_1l_2\dots l_N}|
\approx\sqrt{\frac{m}{(2\pi)^{N-1}l_1l_2\dots l_N}}
\left(\frac{m^{1-\frac{1}{N}}|\kappa_N|^\frac{1}{N}}{e^{1-\frac{1}{N}}}\right)^m
\nonumber\\
&\qquad\qquad
\times\exp\left[\frac{m^{1-\frac{1}{N}}|\kappa_{N-1}|}{(N-1)|\kappa_N|^{1-\frac{1}{N}}}+O(m^{1-\frac{2}{N}})\right].
\label{eq:t206}
\end{align}
For the neighboring terms, one can calculate
\begin{align}
\frac{s_{m;l_1+r_1\dots l_N+r_N}}{s_{m;l_1\dots l_N}}
\approx\prod_{n=1}^{N-1}
 e^{i\Theta_nr_n-\frac{r_n^2}{2l_n}+\dots}\,,
\label{eq:t208}
\end{align}
where $\Theta_n=\psi_n-(n/N)\psi_N$.

Recasting Eq.~(\ref{eq:t202a}) as
\[
Z_m\approx \frac{s_{m;l_1l_2\dots l_N}}{N}\int\limits_{-\infty}^{+\infty}\!\mathrm{d}r_{N-1}\dots \int\limits_{-\infty}^{+\infty}\!\mathrm{d}r_1
 \frac{s_{m;l_1+r_1\dots l_N+r_N}}{s_{m;l_1\dots l_N}}\,,
\]
and evaluating integrals, one finds
\[
Z_m\approx\frac{s_{m;l_1l_2\dots l_N}}{N}\prod_{n=1}^{N-1}\sqrt{2\pi l_n}e^{-\frac{l_n\Theta_n^2}{4}}.
\]

Substituting Eq.~(\ref{eq:t206}), one finally obtains
\begin{align}
&
|Z_m|\approx\frac{1}{\sqrt{N}}\left(\frac{m^{1-\frac{1}{N}}|\kappa_N|^\frac{1}{N}}{e^{1-\frac{1}{N}}}\right)^m
\exp\Bigg[\frac{m^{1-\frac{1}{N}}|\kappa_{N-1}|}{(N-1)|\kappa_N|^{1-\frac{1}{N}}}
\nonumber\\
&\qquad\qquad
\times \left(1-\frac{\Theta_{N-1}^2}{4}\right)
%\nonumber\\
%&\qquad\qquad\qquad
+O(m^{1-\frac{2}{N}})\Bigg].
\label{eq:t209}
\end{align}
Eq.~(\ref{eq:t209}) for $N=2$ is in agreement with Eq.~(\ref{eq:t110}) (by definition, $\Theta_2=\Theta/2$).

According to Eq.~(\ref{eq:t209}), for sufficiently large $m$ (specifically, $m\gg\frac{|\kappa_{N-1}|^N}{|\kappa_N|^{N-1}}+\frac{|\kappa_N|^{N-1}}{|\kappa_{N-1}|^N}$) and $m>e/|\kappa_N|^\frac{1}{N-1}$, one finds $|Z_m|>1$, which is not admitted for $\langle{e^{im\varphi}}\rangle$ of a phase variable $\varphi$.

\section{Comparison to the case of a variable on the line}
\label{sec:2cGauss}
Let us compare the circular cumulant expansions for variables on the circle with the conventional cumulant expansions for variables on the line.

The distribution of a variable $x$ on the line can be fully characterized by its real-valued cumulants $K_m$, $m=1,2,...$, or moments $\mu_m=\langle x^m\rangle$. The case of $K_1\ne0$, $K_{m>1}=0$ corresponds to a $\delta$-function probability density $w(x)=\delta(x-K_1)$ (notice, it is sufficient to assume $K_2=0$, which dictates all $K_{m>2}=0$). The case of two first nonzero cumulants --- the mean value $K_1\ne0$ and the variance $K_2\ne0$ --- and $K_{m>2}=0$ corresponds to the Gaussian distribution
 $w(x)=(2\pi K_2)^{-\frac12}e^{-\frac{(x-K_1)^2}{2K_2}}$.
Thus, one can deal with one- and two-cumulant truncations of the random variable representation. It is important, that no higher order truncations are admitted~\cite{Lukacs-1970}. For instance, assuming $K_3\ne0$ (which enforces as well a nonzero variance $K_2$), one cannot set $K_{m>3}=0$; physically sensible distributions will require at least some higher-order cumulants to be nonzero (even though they can be small and rapidly decay as the order $m$ increases).

For circular cumulants of a phase variable $\varphi$ on the circle, the situation is different; one must have either only one nonzero circular cumulant~\footnote{It is not necessarily that the nonzero cumulant is the first one. Indeed, if the first cumulant $\kappa_1$ is the only nonzero cumulant; transformation $\varphi\to n\varphi$ makes $\kappa_n\ne0$ and $\kappa_{m\ne n}=0$.} or an infinite series of nonzero cumulants. This dissimilarity requires explanations, since the algebraic relations between cumulants $K_m$ and moments $\mu_m$ are equivalent to the algebraic relations between circular cumulants $\kappa_m$ and order parameters $Z_m$ (up to a multiplier, $\kappa_m\leftrightarrow K_m/(m-1)!$\,). We derived that having two nonzero circular cumulants results in large values of high-order $Z_m$, $|Z_m|>1$, while it is not admitted by the physical meaning of $Z_m$. One observes the same situation for the Gaussian distribution with arbitrary nonzero variance $K_2$. Indeed, for $K_2>0$, there is always a finite probability of $|x|>1$; therefore, for large enough $m=2n$, $\langle x^{2n}\rangle>1$ (the case of odd $m$ also yields $|\langle x^{m}\rangle|>1$ if $K_1\ne0$). However, $|\mu_m|>1$ are admitted for a line variable. Thus, the fundamental reasons forbidding finite-number truncations for circular cumulants essentially differ from that for conventional cumulants of line variables.

To summarize, the only admitted truncation for a phase variable is the one-cumulant one and it corresponds to the wrapped Cauchy distribution (or Ott--Antonsen ansatz), while for a line variable, the only admitted nontrivial truncation is the two-cumulant one and it corresponds to the Gaussian distribution.

\begin{figure}[!t]
\center{\footnotesize\sf
(a)\hspace{-12pt}
\includegraphics[width=0.46\columnwidth]%
 {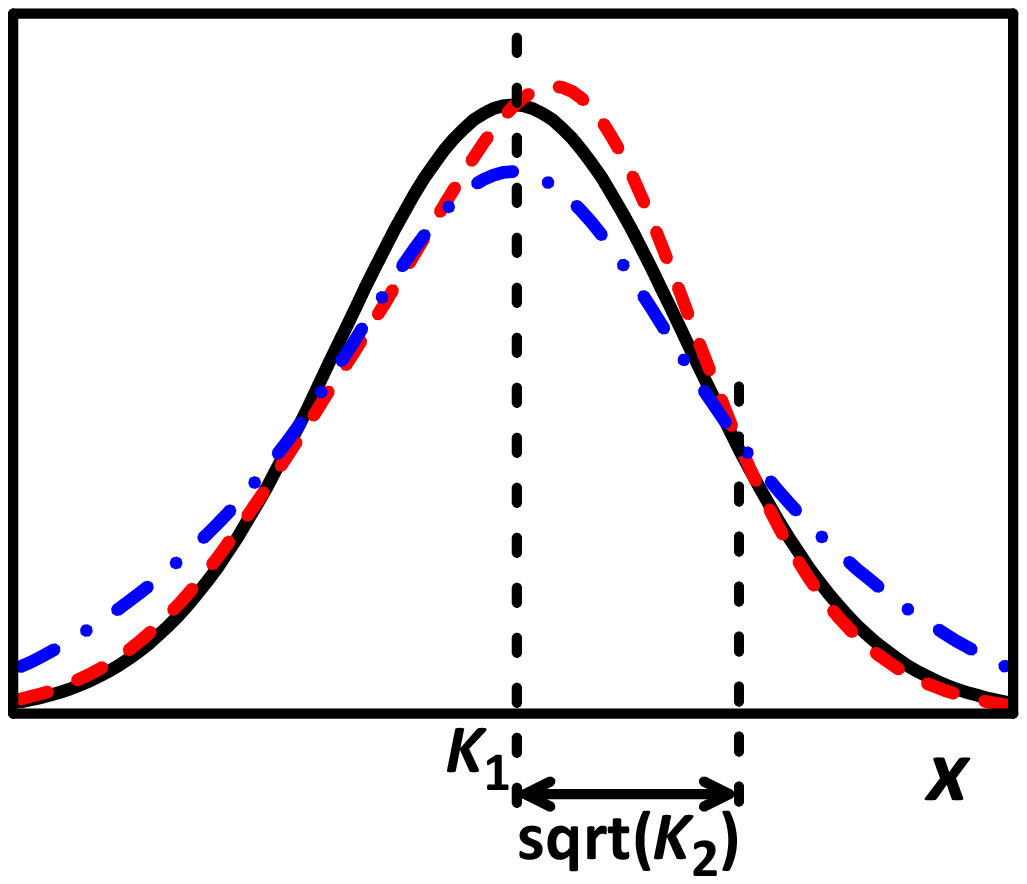}
\quad
(b)\hspace{-12pt}
\includegraphics[width=0.46\columnwidth]%
 {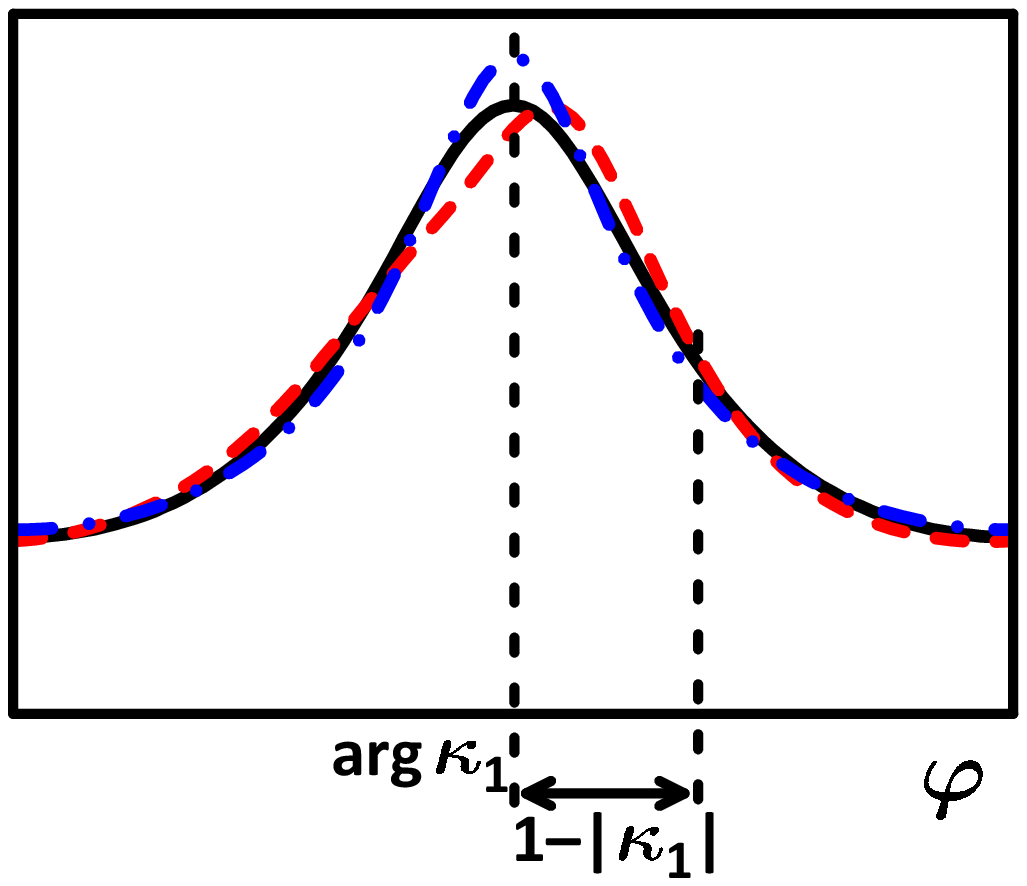}
}
\caption{Black solid lines show the Gaussian (a) and wrapped Cauchy (b) distributions characterized by $\{K_1,K_2\}$ and $\kappa_1$, respectively. Red dashed lines: the perturbed distributions with symmetry braking owned by $K_3\ne0$~(a) and $\arg\kappa_2-2\arg\kappa_1\ne0$~(b). Blue dash-dotted lines: the perturbed distributions without symmetry braking, $K_4\ne0$~(a) and $|\kappa_2|\ne0$, $\arg\kappa_2-2\arg\kappa_1=0$~(b).
}
  \label{fig1}
\end{figure}

Further, we compare the characterization of a distribution by the cumulants for admitted truncations and the meaning of higher-order cumulants. For a line variable, the cumulants are real-valued and two first cumulants quantify the centering of the distribution, $K_1$, and its width $\sqrt{K_2}$ (see Fig.~\ref{fig1}a). The asymmetry of the distribution is quantified by $K_3$ and its kurtosis $K_4$ measures the deviation of tails from the Gaussian law. For a phase variable, the circular cumulants are complex-valued, the argument of the first cumulant $\arg\kappa_1$ features the distribution centering and the absolute value $|\kappa_1|$ features the distribution width (see Fig.~\ref{fig1}b). In particular, $|\kappa_1|=1$ for a $\delta$-function distribution and $|\kappa_1|=0$ for a uniform distribution. The second complex-valued circular cumulant $\kappa_2$ quantifies the distribution asymmetry with $(\arg\kappa_2-2\arg\kappa_1)$ and deformation of tails with $|\kappa_2|$. Thus, in both cases, the reference distribution is characterized by two primary quantities and the principal correction to it has to be characterized by another pair of quantities. However, these quantities involve different number of real- and complex-valued cumulants.

%%%%%%%%%%%%%%%%%%%%%%%%%%%%%%%%%%%%%%%%%%%%%%%%%%%%%%%%%%%%
%%%%%%%%%%%%%%%%%%%%%%%%%%%%%%%%%%%%%%%%%%%%%%%%%%%%%%%%%%%%

\section{Macroscopic variables for populations of quadratic integrate-and-fire neurons}
\label{sec:QIF}
While a plain discarding of higher-order cumulants in cumulant equation chains can frequently be a reasonable approximation, one can encounter situations which require a more subtle treatment. In these situations, a formal adopting of finite cumulant truncations can lead to the divergence of the mean fields mediating the interaction between population elements. Below we consider an example of such a system.

The population of quadratic integrate-and-fire neurons (QIFs) (e.g.,~\cite{Montbrio-Pazo-Roxin-2015,Pazo-Montbrio-2016,Volo-Torcini-2018}) obeys
\begin{align}
&\dot{V}_j=V_j^2+I_j\,,
\label{eq:t301}
\\
&I_j=\eta_j+Js(t)+I(t)\,,
\label{eq:t302}
\end{align}
where $V_j$ and $I_j$ represent a neuron's membrane potential and an input current, respectively, $\eta_j$ and $I(t)$ are individual and common parts of the input current, respectively, $s(t)$ is a common field proportional to the firing rate $r(t)$, and $J$ is the synaptic weight. One can introduce a phase variable $\varphi$,
\begin{equation}
V_j=\tan\frac{\varphi_j}{2},
\label{eq:V-phi}
\end{equation}
and rewrite Eq.~(\ref{eq:t301}) in its terms:
\begin{equation}
\dot\varphi_j=(1-\cos\varphi_j)+(1+\cos\varphi_j)\left[\eta_j+Js(t)+I(t)\right].
\nonumber
\end{equation}

For this system the variable
\begin{equation}
W(t)=1-2Z_1+2Z_2-2Z_3+2Z_4-\dots
\label{eq:t303}
\end{equation}
%where $Z_m=\langle{e^{im\varphi}}\rangle$,
is important.
Indeed, one can show~\cite{Montbrio-Pazo-Roxin-2015} that the voltage mean-field
\begin{equation}
v=\langle V\rangle=\mathrm{p.v.}\int_{-\pi}^{\pi}\mathrm{d}\varphi\tan\frac{\varphi}{2}\,w(\varphi)=-\mathrm{Im}(W)
\nonumber %\label{eq:t304}
\end{equation}
and the firing rate $r(t)$ in terms of $\varphi$
%for a quasi stationary probability density $w(\varphi)$ (the mean-field dynamics should be slow as compared to the reference firing rate of a single neuron; for the mean voltage the condition of quasistationarity is not required)
is
\begin{equation}
r(t)=2w(\pi)=\mathrm{Re}(W)/\pi\,.
\nonumber %\label{eq:t305}
\end{equation}
Thus,
\begin{equation}
W(t)=\pi r(t)-iv(t)\,;
\label{eq:t304}
\end{equation}
the population dynamics is essentially controlled by $s(t)\propto r(t)=\mathrm{Re}(W)/\pi$ and $\mathrm{Im}(W)$ is an important macroscopic observable. The variable $r(t)$ is generally important for populations of pulse-coupled oscillators~\cite{Pazo-Montbrio-2014,Montbrio-Pazo-Roxin-2015,Pazo-Montbrio-2016,Devalle-Pazo-Montbrio-2018,Volo-Torcini-2018}.

On the Ott--Antonsen manifold $Z_m=(Z_1)^m$, the sum in (\ref{eq:t303}) can be readily calculated:
\[
W_\mathrm{OA}=\frac{1-Z}{1+Z}\,.
\]

\subsection{Divergence of firing rate for any two-cumulant truncation}
\label{ssec:2cQIF}
If only two first cumulants $\kappa_1$ and $\kappa_2$ are nonzero, the circular cumulant generating function $\Psi(k)=\kappa_1k+\kappa_2k^2$, and, see Eq.~(\ref{eq:t101}), the moment generating function
\begin{align}
&\textstyle
F
%=e^{\kappa_1k}e^{\kappa_2\frac{k^2}{2}}
%\nonumber\\
%&\quad\textstyle
%{}=e^{\kappa_1k}\big(1+\frac{\kappa_2}{2}k^2+\frac{\kappa_2^2}{8}k^4 +\dots\big)
%\nonumber\\
%&\quad\textstyle
%{}
=\sum\limits_{m=0}^{\infty}\kappa_1^m\frac{k^m}{m!}\big(1+\frac{\kappa_2}{2}k^2+\frac{\kappa_2^2}{8}k^4 +\dots\big)\,.
\nonumber
%\label{eq:t305}
\end{align}
Since $\sum_{m=0}^{\infty}\kappa_2\kappa_1^m\frac{k^{m+2}}{m!} =\sum_{m=0}^{\infty}m(m-1)\kappa_2\kappa_1^{m-2}\frac{k^m}{m!}$, etc., one finds
\begin{align}
&\textstyle
Z_m=\kappa_1^m+\frac{m(m-1)}{2}\kappa_2\kappa_1^{m-2}
\nonumber\\
&\qquad\qquad\textstyle
{}+\frac{m(m-1)(m-2)(m-3)}{8}\kappa_2^2\kappa_1^{m-4}+\dots
\nonumber
%\label{eq:t306a}
\\
&\quad\textstyle
{}=\big(1+\frac{\kappa_2}{2}\frac{\partial^2}{\partial\kappa_1^2} +\frac{\kappa_2^2}{8}\frac{\partial^4}{\partial\kappa_1^4}+\dots\big)\kappa_1^m
\nonumber
%\label{eq:t306b}
\\
&\quad\textstyle
{}=\mathrm{exp}(\frac{\kappa_2}{2}\frac{\partial^2}{\partial\kappa_1^2}\big)\kappa_1^m\,.
\nonumber
%\label{eq:t306c}
\end{align}
Here, the exponential $\mathrm{exp}(\hat{Q})$ of an operator $\hat{Q}$ is defined as $1+\hat{Q}+\frac12\hat{Q}^2+\frac{1}{3!}\hat{Q}^3+\dots$\,.
Hence,
\begin{align}
&\textstyle
W=1-2Z_1+2Z_2-2Z_3+\dots=
\nonumber\\
&\quad\textstyle
{}=\mathrm{exp}(\frac{\kappa_2}{2}\frac{\partial^2}{\partial\kappa_1^2}\big)
(1-2\kappa_1+2\kappa_1^2-2\kappa_1^3+\dots)
\nonumber\\
%&\quad\textstyle
%{}=\mathrm{exp}(\frac{\kappa_2}{2}\frac{\partial^2}{\partial\kappa_1^2}\big)
% \big(\frac{2}{1+\kappa_1}-1\big)
%\nonumber\\
&\quad\textstyle
=2\mathrm{exp}(\frac{\kappa_2}{2}\frac{\partial^2}{\partial\kappa_1^2}\big)
 \frac{1}{1+\kappa_1} -1\,.
\label{eq:t307}
\end{align}
One can calculate
\begin{equation}
\frac{\partial^n}{\partial\kappa_1^n}\frac{1}{1+\kappa_1}=\frac{(-1)^nn!}{(1+\kappa_1)^{n+1}},
\label{eq:t308}
\end{equation}
and Eq.~(\ref{eq:t307}) yields a series
\begin{align}
&\textstyle
W=2\sum_{m=0}^\infty\frac{1}{m!}\frac{\kappa_2^m}{2^m}\frac{\partial^{2m}}{\partial\kappa_1^{2m}}
 \frac{1}{1+\kappa_1} -1
\nonumber\\
%&\quad\textstyle
%=\frac{1-\kappa_1}{1+\kappa_1}+2\sum_{m=1}^\infty\frac{1}{m!}\frac{\kappa_2^m}{2^m} %\frac{(2m)!}{(1+\kappa_1)^{2m+1}}
%\nonumber\\
&\quad\textstyle
=\frac{1-\kappa_1}{1+\kappa_1}+\frac{2}{1+\kappa_1}\sum_{m=1}^\infty (2m-1)!!\big[\frac{\kappa_2}{(1+\kappa_1)^2}\big]^m,
\label{eq:t309}
\end{align}
where $(2m-1)!!\equiv1\cdot...\cdot(2m-3)\cdot(2m-1)$\,.

The series (\ref{eq:t309}) diverges for arbitrary $\kappa_2$. Indeed, for $m>\frac{|1+\kappa_1|^2}{2|\kappa_2|}$, the sum terms grow with $m$.

\subsection{Untruncated circular cumulant expansions}
\label{ssec:NcQIF}
Now we rewrite $W$ [Eq.~(\ref{eq:t303})] in terms of circular cumulants for untruncated expansions. The circular cumulant generating function $\Psi(k)=\kappa_1k+\kappa_2k^2+\kappa_3k^3+...$\,, and
\begin{align}
&\textstyle
F=e^{\kappa_1k}e^{\kappa_2\frac{k^2}{2}+\kappa_3\frac{k^3}{3}+...}
\nonumber\\
&\quad\textstyle
{}=e^{\kappa_1k}\big(1+\frac{\kappa_2}{2}k^2+\frac{\kappa_3}{3}k^3+\frac{\kappa_2^2}{8}k^4 +\dots\big)
\nonumber\\
&\quad\textstyle
{}=\sum\limits_{m=0}^{\infty}\kappa_1^m\frac{k^m}{m!}\big(1+\frac{\kappa_2}{2}k^2+\frac{\kappa_3}{3}k^3+\frac{\kappa_2^2}{8}k^4 +\dots\big)\,.
\nonumber
%\label{eq:t310}
\end{align}
Since $\sum_{m=0}^{\infty}\kappa_2\kappa_1^m\frac{k^{m+2}}{m!} =\sum_{m=0}^{\infty}m(m-1)\kappa_2\kappa_1^{m-2}\frac{k^m}{m!}$, etc., one finds
\begin{align}
&\textstyle
Z_m=\kappa_1^m+\frac{m(m-1)}{2}\kappa_2\kappa_1^{m-2} +\frac{m(m-1)(m-2)}{3}\kappa_3\kappa_1^{m-3}
\nonumber\\
&\qquad\qquad\textstyle
{}+\frac{m(m-1)(m-2)(m-3)}{8}\kappa_2^2\kappa_1^{m-4}+\dots
\nonumber\\
&\quad\textstyle
{}=\big(1+\frac{\kappa_2}{2}\frac{\partial^2}{\partial\kappa_1^2} +\frac{\kappa_3}{3}\frac{\partial^3}{\partial\kappa_1^3} +\frac{\kappa_2^2}{8}\frac{\partial^4}{\partial\kappa_1^4}+\dots\big)\kappa_1^m
\nonumber
%\label{eq:t311a}
\\
&\quad\textstyle
{}=\mathrm{exp}(\frac{\kappa_2}{2}\frac{\partial^2}{\partial\kappa_1^2} +\frac{\kappa_3}{3}\frac{\partial^3}{\partial\kappa_1^3}+\dots\big)\kappa_1^m\,.
\nonumber
%\label{eq:t311}
\end{align}
Hence,
\begin{align}
&\textstyle
W=1-2Z_1+2Z_2-2Z_3+\dots=
\nonumber\\
&\quad\textstyle
{}=\mathrm{exp}(\frac{\kappa_2}{2}\frac{\partial^2}{\partial\kappa_1^2} +\frac{\kappa_3}{3}\frac{\partial^3}{\partial\kappa_1^3}+\dots\big)
\nonumber\\
&\qquad\qquad\textstyle
\times(1-2\kappa_1+2\kappa_1^2-2\kappa_1^3+\dots)
\nonumber\\
%&\quad\textstyle
%{}=\mathrm{exp}(\frac{\kappa_2}{2}\frac{\partial^2}{\partial\kappa_1^2} +\frac{\kappa_3}{3}\frac{\partial^3}{\partial\kappa_1^3}+\dots\big)
% \big(\frac{2}{1+\kappa_1}-1\big)
%\nonumber\\
&\quad\textstyle
=2\mathrm{exp}(\frac{\kappa_2}{2}\frac{\partial^2}{\partial\kappa_1^2} +\frac{\kappa_3}{3}\frac{\partial^3}{\partial\kappa_1^3}+\dots\big)\frac{1}{1+\kappa_1}
 -1\,,
\label{eq:t312}
\end{align}
where one can use formula~(\ref{eq:t308}).
%One can calculate
%$\frac{\partial^n}{\partial\kappa_1^n}\frac{1}{1+\kappa_1}=\frac{(-1)^nn!}{(1+\kappa_1)^{n+1}}$.

\subsection{Finite-$N$ cumulant approximations}
\label{ssec:NcApp}
In this section we address the question whether one can use finite number cumulant approximations in applications. For approximations a small parameter is required. Refs.~\cite{Tyulkina-etal-2018,Goldobin-etal-2018,Goldobin-2019,Goldobin-Dolmatova-2019b} theoretically reveal the importance and persistence of the case where circular cumulants obey hierarchy $\kappa_n\propto\varepsilon^{n-1}$ with a small number $\varepsilon$. Below, in Sec.~\ref{sec:hrch}, we will discuss this hierarchy and report it to be highly relevant for experimental data as well.

The hierarchy $\kappa_n\propto\varepsilon^{n-1}$ is essentially important for a rigorous approach to constructing approximations. One should construct an expansion with respect to a small parameter $\varepsilon$, which means that if one introduces $\kappa_2^2$-corrections, then $\kappa_3$ should be also taken into account for to achieve the accuracy $o(\kappa_2^2)$, etc. In particular, up to the first-order corrections, Eq.~(\ref{eq:t312}) yields
\begin{equation}
W=\frac{1-\kappa_1}{1+\kappa_1}+\frac{2\kappa_2}{(1+\kappa_1)^3} +O(\varepsilon^2)\,;
\label{eq:t313}
\end{equation}
up to the second-order corrections,
\begin{equation}
W=\frac{1-\kappa_1}{1+\kappa_1}+\frac{2\kappa_2}{(1+\kappa_1)^3} -\frac{4\kappa_3}{(1+\kappa_1)^4} +\frac{6\kappa_2^2}{(1+\kappa_1)^5}+O(\varepsilon^3)\,.
\label{eq:t314}
\end{equation}

Quite often in applications, one adopts a certain approximation instead of constructing a rigorous expansion. Such an approximation has a rigorously guaranteed order of accuracy. Additionally, the approximation can not only ease analytical calculations but also effectively yield a much higher accuracy than the rigorously guaranteed one. In the case of circular cumulant representation, one should be careful with such approximations; while one can introduce finite number of corrections related to higher cumulants, one cannot adopt approximations which correspond to exact formal expressions for a finite number of nonzero circular cumulants. For instance, in Sec.~\ref{ssec:2cQIF}, for two nonzero cumulants, the macroscopic variable $W$ exactly corresponding to these cumulants diverges for arbitrary nonzero $\kappa_2$. In detail, the first correction with the $\kappa_2$-term is accurate; the further $\kappa_2^n$-contributions for moderate $n$ introduce smaller corrections which do not enhance the accuracy for a specific distribution $w(\varphi)$; for large $n$, these excessive corrections start to diverge.

%%%%%%%%%%%%%%%%%%%%%%%%%%%%%%%%%%%%%%%%%%%%%%%%%%%%%%%%%%%%
\begin{figure}[!t]
\center{\footnotesize\sf
%(a)\hspace{-12pt}
\includegraphics[width=0.48\columnwidth]%
 {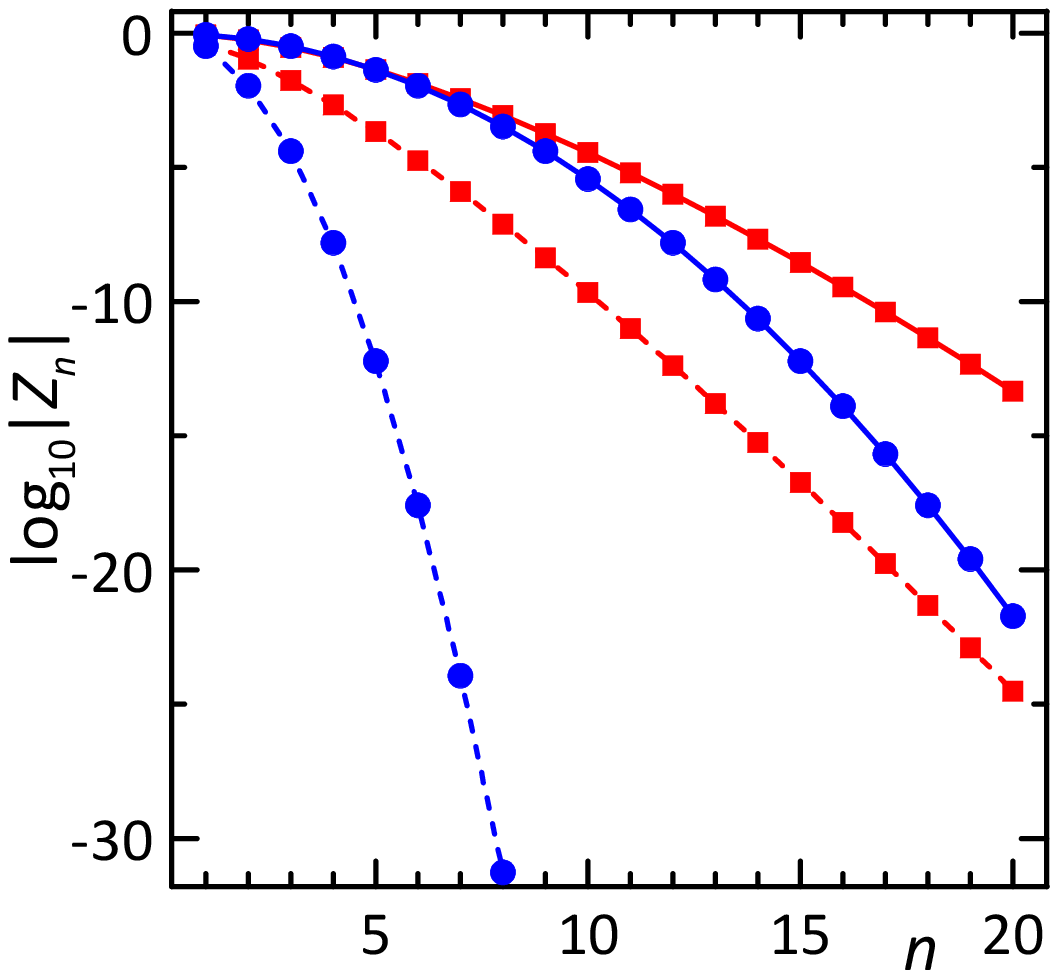}
\;
%(b)\hspace{-12pt}
\includegraphics[width=0.48\columnwidth]%
 {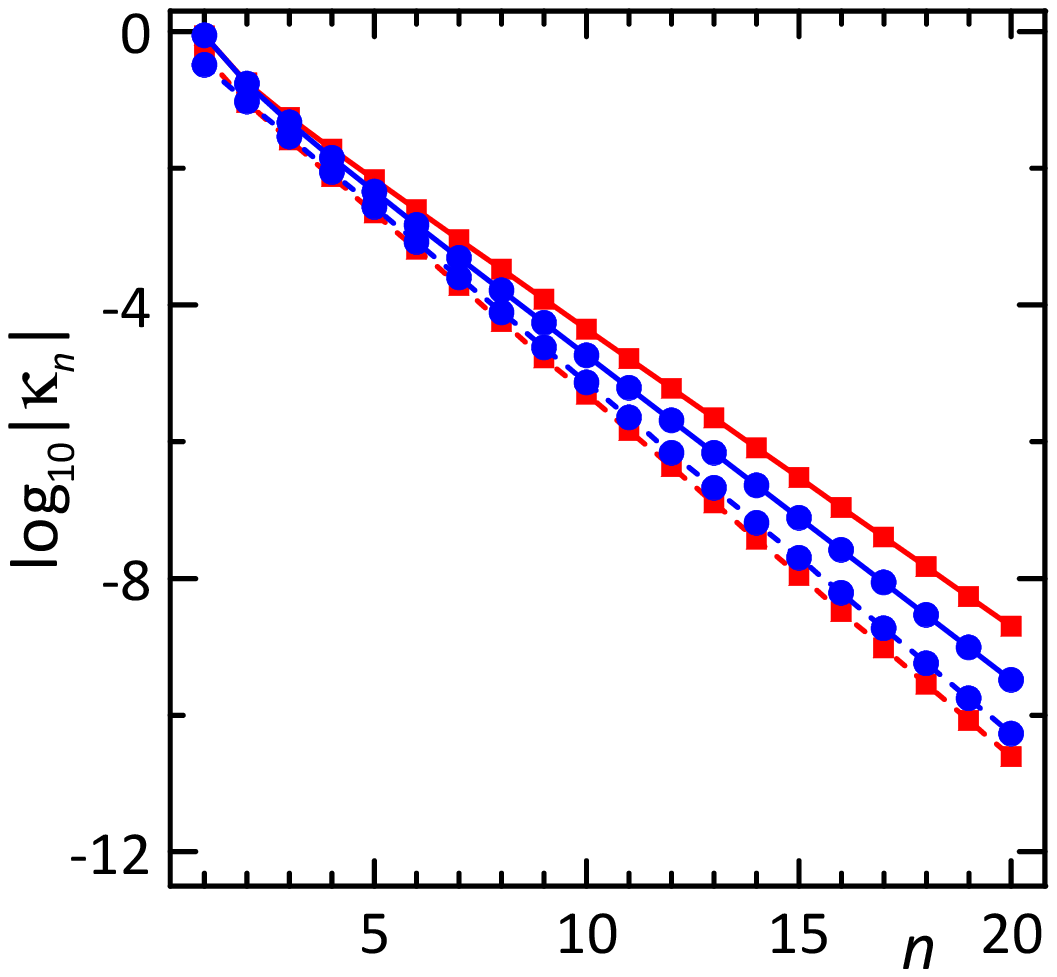}
}
\caption{Order parameters $Z_n$ (left panel) and circular cumulants $\kappa_n$ (right panel) are plotted for the wrapped Gaussian distribution~(\ref{eq:wG}) with blue circles (solid line: $\sigma=0.5$, dashed line: $\sigma=1.5$) and for the von Mises distribution~(\ref{eq:vM}) with red squares (solid line: $\sigma=0.5$, dashed line: $\sigma=1$).
}
  \label{fig2}
\end{figure}
%%%%%%%%%%%%%%%%%%%%%%%%%%%%%%%%%%%%%%%%%%%%%%%%%%%%%%%%%%%%

\section{Circular cumulant hierarchies}
\label{sec:hrch}
\subsection{Wrapped Gaussian and von Mises distributions}
\label{ssec:wGvM}
Let us consider two important particular distributions:
\\
(i)~wrapped Gaussian distribution
\begin{equation}
w_\mathrm{wG}(\varphi)=\sum_{n=-\infty}^{+\infty}\frac{1}{\sqrt{2\pi}\sigma}
 e^{-\frac{(\varphi-\psi+2\pi n)^2}{2\sigma^2}}
\label{eq:wG}
\end{equation}
of half-width $\sigma$ around $\psi$, which is relevant for some phase ensembles~\cite{Zaks-etal-2003,Sonnenschein-Schimansky-Geier-2013,Sonnenschein-etal-2015,Hannay-Forger-Booth-2018};
\\
(ii)~von Mises distribution
\begin{equation}
w_\mathrm{vM}(\varphi)=\frac{e^{\frac{\cos(\varphi-\psi)}{\sigma^2}}}{2\pi I_0(\sigma^{-2})}
\label{eq:vM}
\end{equation}
of half-width $\sigma$ around $\psi$, here $I_n(\cdot)$ is the $n$-th order modified Bessel function of the first kind. Von Mises distribution features the steady states of ensembles of identical phase elements subject to additive intrinsic noise and common static field (e.g., see~\cite{Bertini-Giacomin-Pakdaman-2010} or \cite{Goldobin-etal-2018}).

With these distributions one can calculate the macroscopic variable $W$. For wrapped Gaussian distribution~(\ref{eq:wG}), $Z_n=e^{-\sigma^2n^2/2}e^{in\psi}$, and series~(\ref{eq:t303}) obviously possesses good convergence properties:
\begin{equation}
W_\mathrm{wG}=
 1+2\sum_{n=1}^{+\infty}e^{in(\psi+\pi)}e^{-\frac{1}{2}\sigma^2n^2}.
\label{eq:WwG}
\end{equation}
For von Mises distribution~(\ref{eq:vM}), with the Jacobi--Anger expansion $e^{a\cos(\varphi-\psi)}=\sum_{n=-\infty}^{+\infty}I_n(a)e^{in(\varphi-\psi)}$, one finds $Z_n=[I_n(\sigma^{-2})/I_0(\sigma^{-2})]e^{in\psi}$, and series~(\ref{eq:t303}) reads
\begin{equation}
W_\mathrm{vM}=
 1+2\sum_{n=1}^{+\infty}e^{in(\psi+\pi)}\frac{I_n(\sigma^{-2})}{I_0(\sigma^{-2})}\,.
\label{eq:WvM}
\end{equation}

In Fig.~\ref{fig2}, one can see, that series $|Z_n|$ decay fast for both distributions providing a good convergence of $W$. Simultaneously, circular cumulants form a clearly pronounced geometric progression even for moderate values of parameter $\sigma$. Thus, for these two distributions the macroscopic variable $W$ obviously converges, while the calculations with a finite number of cumulants has to be performed as discussed in Sec.~\ref{ssec:NcApp}.
In Appendix~\ref{asec:r}, the formulae for $W$~(\ref{eq:WwG}), (\ref{eq:WvM}), (\ref{eq:WwnC}) are confirmed with an alternative approach to calculation of $r$.

%%%%%%%%%%%%%%%%%%%%%%%%%%%%%%%%%%%%%%%%%%%%%%%%%%%%%%%%%%%%
\begin{figure}[!t]
\center{\footnotesize\sf
%(a)\hspace{-12pt}
\includegraphics[width=0.48\columnwidth]%
 {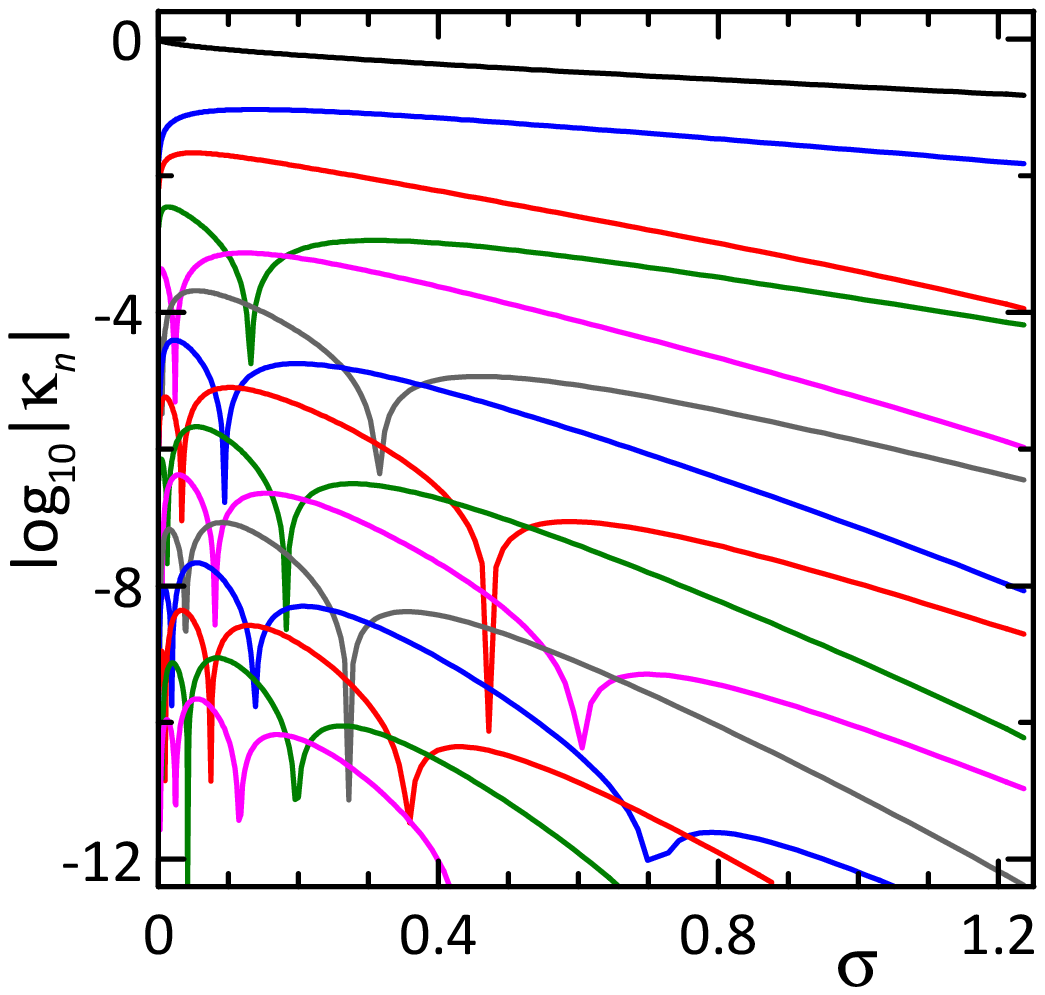}
\;
%(b)\hspace{-13pt}
\includegraphics[width=0.48\columnwidth]%
 {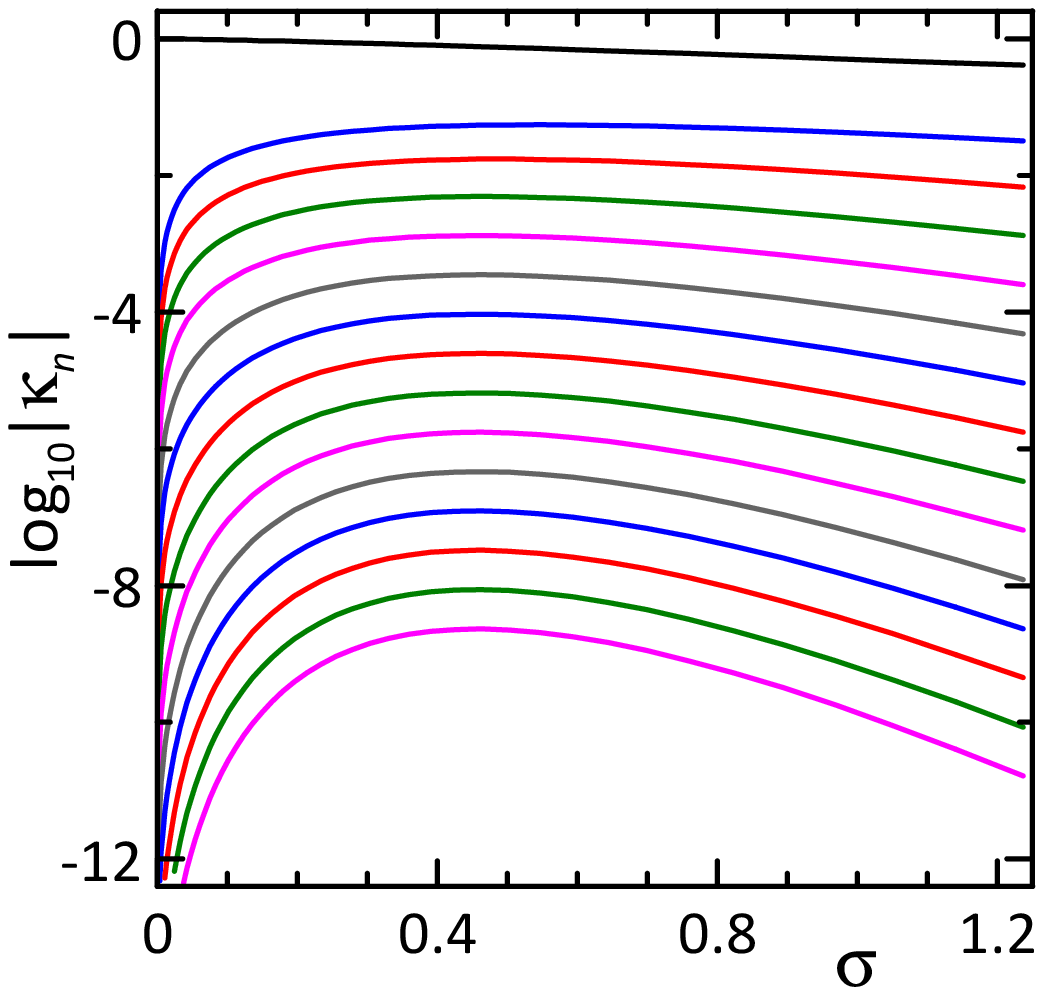}
}
\caption{Circular cumulants $\kappa_n$ are plotted vs distribution half-width $\sigma$ for the wrapped non-Cauchy heavy-tail distribution~(\ref{eq:wnC}) for parameter $\mu=-0.25$ (left panel) and $+0.25$ (right panel); cumulant order $n$ changes from $1$ to $15$ (from top to bottom on the right hand side of panels).
}
  \label{fig3}
\end{figure}
%%%%%%%%%%%%%%%%%%%%%%%%%%%%%%%%%%%%%%%%%%%%%%%%%%%%%%%%%%%%

\subsection{Wrapped non-Cauchy distributions with heavy tails}
\label{ssec:wnC}
The phenomenon of synchronization by common noise is an important case where the Cauchy distribution forms in populations of general limit-cycle oscillators subject to a weak intrinsic noise additionally to the common noise driving (cf Eq.~(16) in~\cite{Goldobin-Pikovsky-2005}). Noticeably, the distribution of phase deviations in high-synchrony regimes is Cauchy even though both the common and intrinsic noises are Gaussian. When the common-noise synchronization mechanism is affected by global coupling, the distribution changes to
\[
w(\theta)=\frac{\Gamma(1+\mu)}{\sqrt{\pi}\sigma\Gamma(\frac12+\mu)} \left(1+\frac{\theta^2}{\sigma^2}\right)^{-(1+\mu)}
\]
(cf Eq.~(18) in~\cite{Goldobin-Dolmatova-2019}), where $\theta$ is the phase deviation from the cluster center; the distribution half-width $\sigma$ is proportional to the intrinsic noise strength and $\sigma\propto(-\lambda)^{-1/2}$, $\lambda$ is the Lyapunov exponent of an oscillator without intrinsic noise; $\mu=[\mbox{coupling strength}]/(-2\lambda)$. Perfect synchrony of identical oscillators without intrinsic noise occurs for $\mu>-1/2$.

For non-high synchrony, non-small deviations of $\theta$ not only make the distribution $w(\theta)$ wrapped within the interval $(-\pi,\pi]$, but also affect the distribution shape due to nonlinearities. Nonetheless, one can consider wrapped non-Cauchy distributions
\begin{equation}
w_\mathrm{wnC}(\theta)=\sum_{n=-\infty}^{+\infty}\frac{\Gamma(1+\mu)}{\sqrt{\pi}\sigma\Gamma(\frac12+\mu) \left(1+\frac{(\theta+2\pi n)^2}{\sigma^2}\right)^{1+\mu}}
\label{eq:wnC}
\end{equation}
as generic ones for synchronization by common noise where it interplays with the coupling between oscillators.

For wrapped non-Cauchy distribution~(\ref{eq:wnC}) one can calculate order parameters
\[
Z_n=\frac{(n\sigma)^{\frac12+\mu}\,\mathrm{K}_{\frac12+\mu}(n\sigma)}{2^{-\frac12+\mu}\,\Gamma(\frac12+\mu)}\,,
\]
where $\mathrm{K}_n(\cdot)$ is the modified second-kind Bessel function. For $\mu>-1/2$, $Z_n$ decay with $n$ exponentially (in detail, for $n\sigma\gg1+\mu^2$, $(n\sigma)^{\frac12+\mu}\mathrm{K}_{\frac12+\mu}(n\sigma)\approx\sqrt{\pi/2}(n\sigma)^\mu e^{-n\sigma}$), and the sum~(\ref{eq:t303})
\begin{equation}
W_\mathrm{wnC}=1+\sum_{n=1}^{+\infty} \frac{4(-1)^n(\frac{n\sigma}{2})^{\frac12+\mu}\,\mathrm{K}_{\frac12+\mu}(n\sigma)}{\Gamma(\frac12+\mu)}
\label{eq:WwnC}
\end{equation}
possesses good convergence properties.

Circular cumulants can be now calculated from $Z_n$. For $\mu>0$, cumulants always form a geometric progression (the right panel in Fig.~\ref{fig3}); for $-1/2<\mu<0$, the geometric progression is somewhat distorted by passings of cumulants through zero, where they change their signs (cusps in the left panel of Fig.~\ref{fig3}), but their reference order of magnitude obeys the same geometric progression hierarchy. The calculations with a finite number of cumulants have to be performed as discussed in Sec.~\ref{ssec:NcApp}.

%%%%%%%%%%%%%%%%%%%%%%%%%%%%%%%%%%%%%%%%%%%%%%%%%%%%%%%%%%%%
\begin{figure}[!t]
\center{\footnotesize\sf
(a)\hspace{-14pt}
\includegraphics[width=0.875\columnwidth]%
 {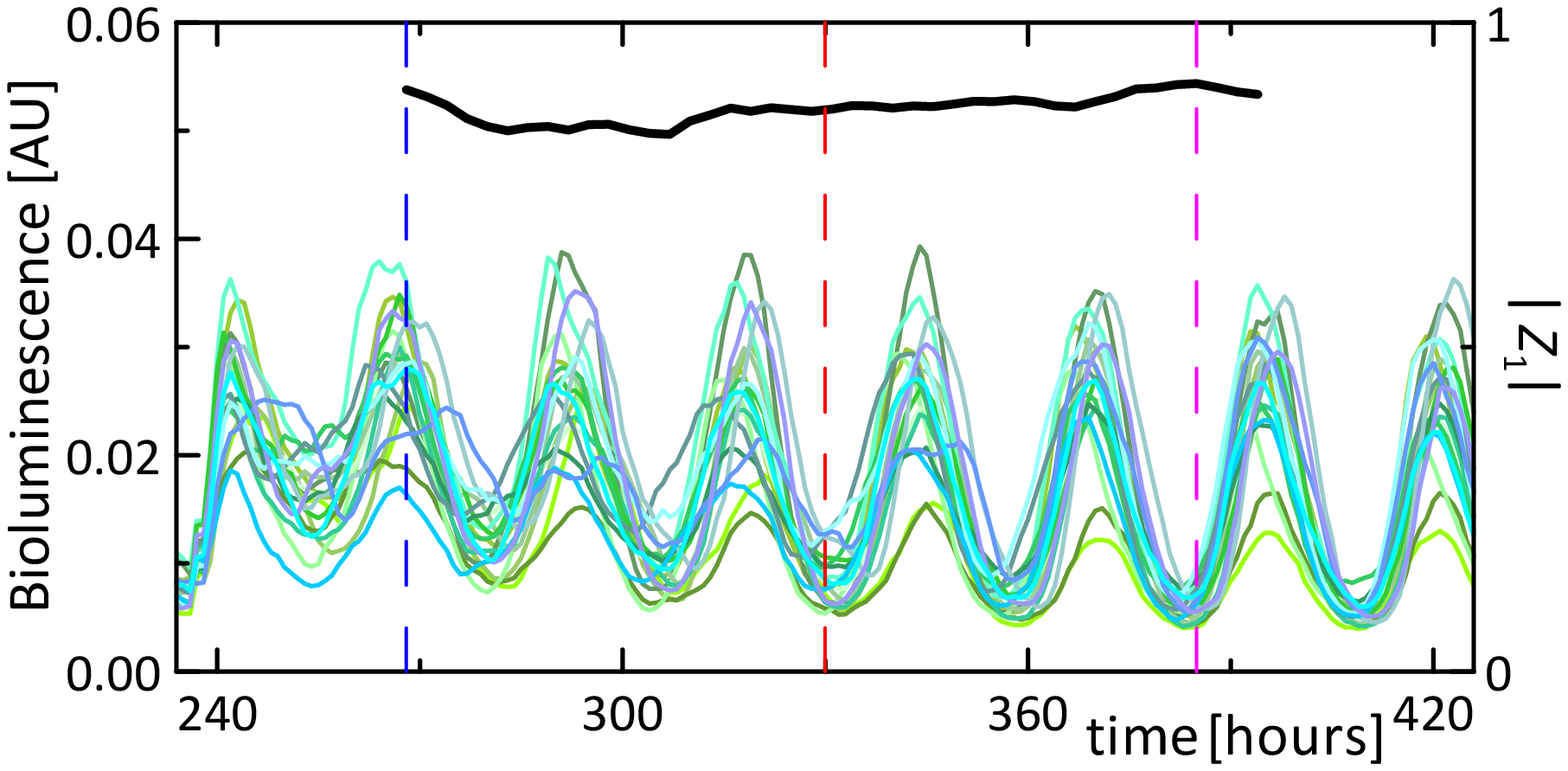}
\\[5pt]
(b)\hspace{-14pt}
\includegraphics[width=0.875\columnwidth]%
 {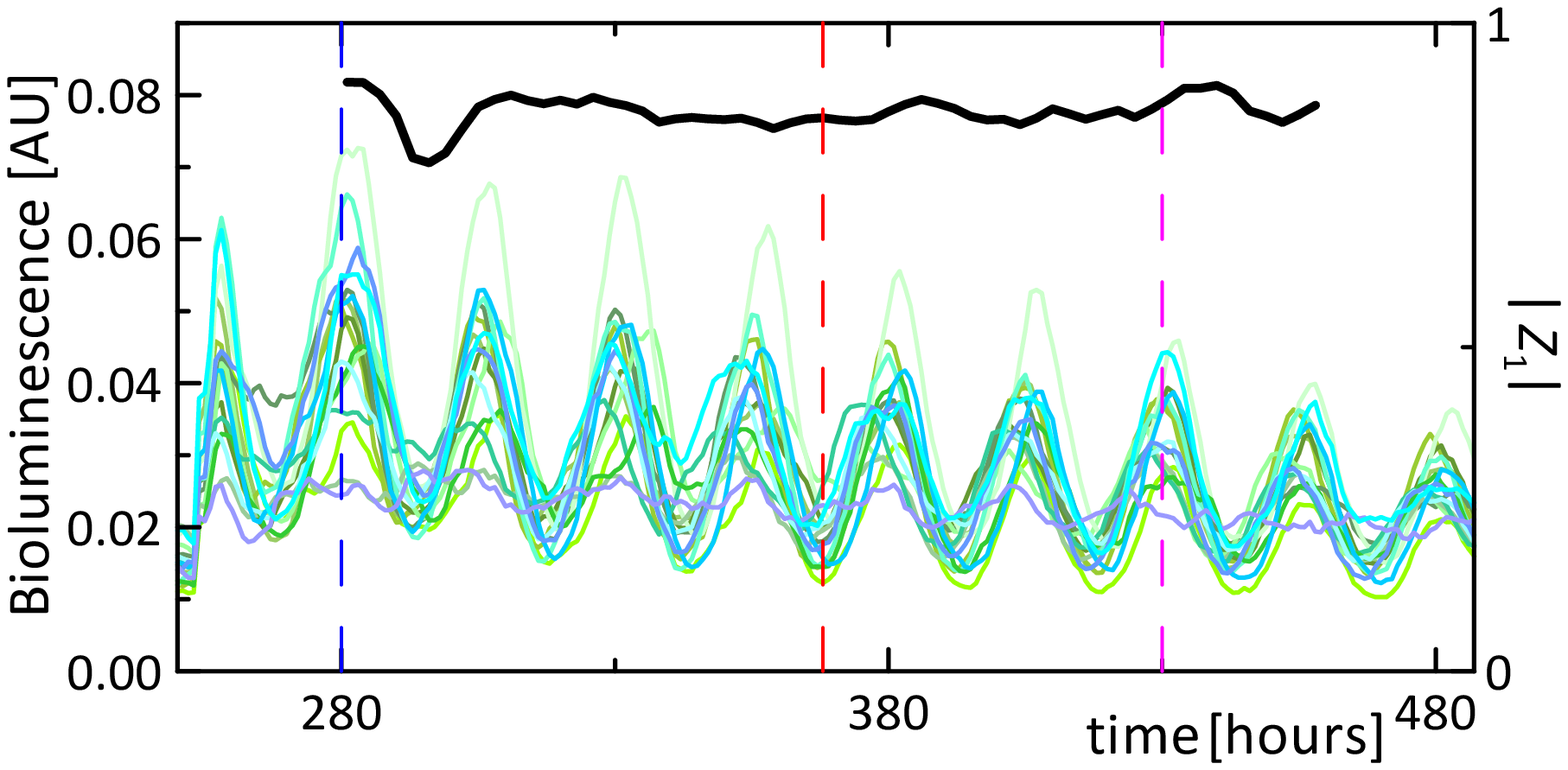}
\\[9pt]
(c)\hspace{-14pt}
\includegraphics[width=0.483\columnwidth]%
 {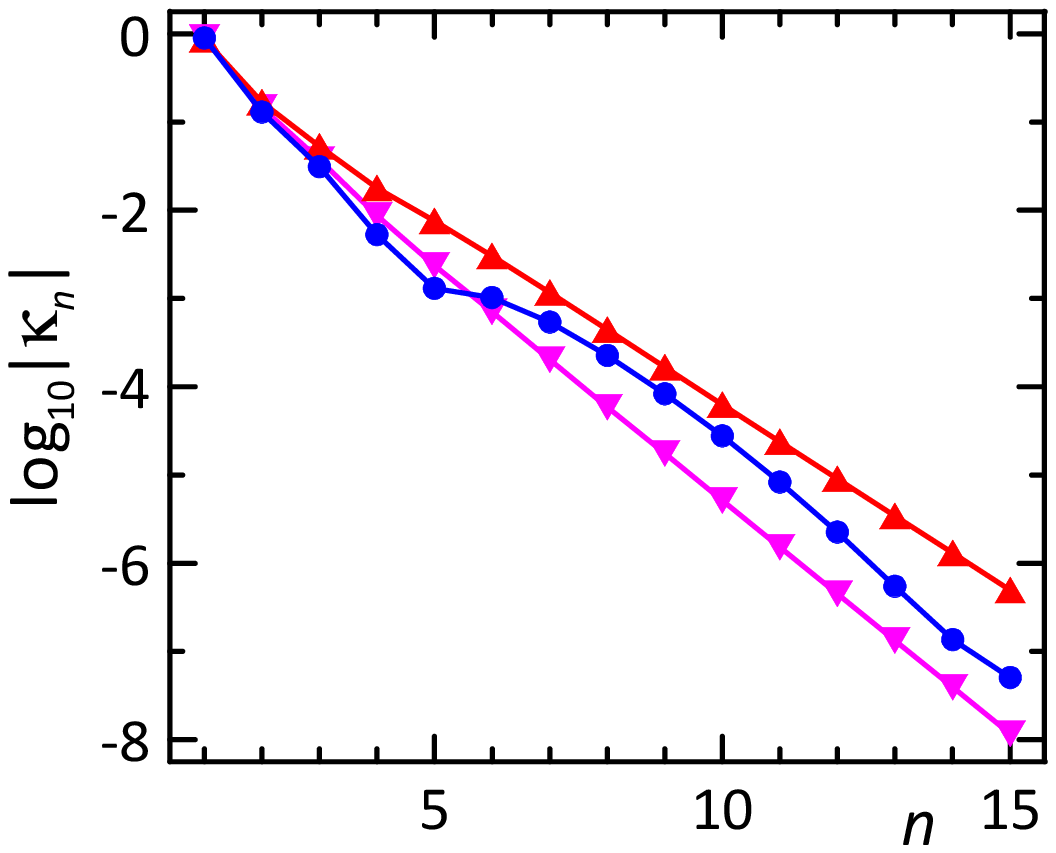}
\;
(d)\hspace{-14pt}
\includegraphics[width=0.483\columnwidth]%
 {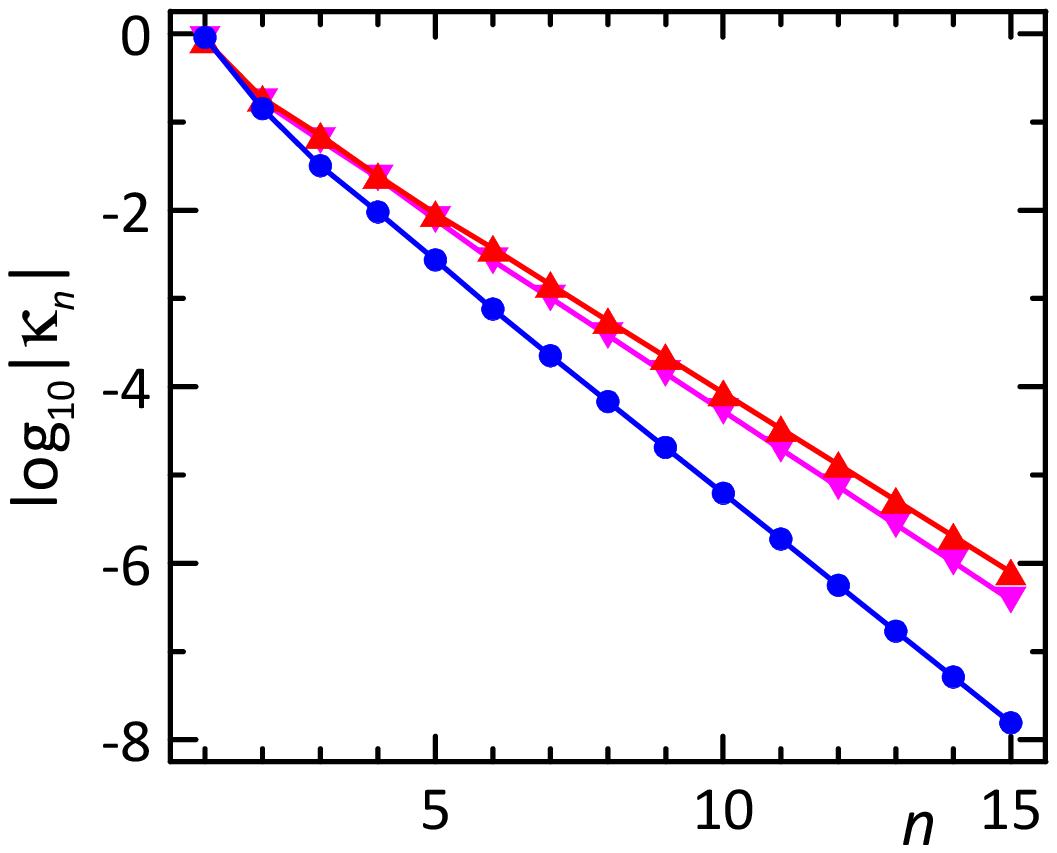}
}
\caption{The order parameter $|Z_1|$ is plotted with the bold black lines for 383 SCN1 cells in panel~(a) and for 264 SCN2 cells in panel~(b); light-color lines show the bioluminescence raw data for 20 arbitrary-chosen cells~\cite{Abel-etal-2016}. (c):~The first 15 circular cumulants for SCN1 at $t=268$ (blue circles), $330$ (red up-pointed triangles), and $385$ (magenta down-pointed triangles); the time instants are marked with vertical dashed lines in panel~(a). (d):~The first 15 circular cumulants for SCN2 at $t=280$ (blue circles), $368$ (red up-pointed triangles), and $430$ (magenta down-pointed triangles); the time instants are marked with dashed lines in panel~(b).
}
  \label{fig4}
\end{figure}
%%%%%%%%%%%%%%%%%%%%%%%%%%%%%%%%%%%%%%%%%%%%%%%%%%%%%%%%%%%%

\subsection{Networks of coupled biological oscillators}
\label{ssec:SCN}
Abel et al.~\cite{Abel-etal-2016} report hourly experimental data on resynchronization of cells in several biologically distinct mammalian suprachiasmatic nucleus (SCN) explants; measurements are performed with single-cell resolution. For illustration, we analyze the bioluminescence oscillation data of individual cells after application of tetrodotoxin for temporary inhibition of intercellular couplings.
In Fig.~\ref{fig4}, the results of analysis of the experimental data for the first (SCN1) and second (SCN2) data sets (publicly available online at https://github.com/JohnAbel/scn-resynchronization-data-2016) from~\cite{Abel-etal-2016} are presented. Protophase is calculated via the Hilbert transform of the individual cell signal. The phases are calculated on the basis of the assumption of an identical functional relation between the protophase and the phase for all cells. The distribution of phases of all oscillators averaged over the integer number of revolutions should be uniform; hence, the corresponding distribution of protophases yields the functional relation between the protophase and the genuine phase~\cite{Kralemann-etal-2007}.

One can see that the circular cumulants form geometric progressions and decay quite rapidly with  order $n$. During transitions between different regimes of collective behavior, a defect of the progression multiplier propagates along the series from low- to high-order cumulants (see the series for SNC1 at $t=268$ in Fig.~\ref{fig4}c).

\subsection{Networks of coupled electrochemical oscillators}
\label{ssec:ElCh}
\begin{figure}[!t]
\center{\footnotesize\sf
(a)\hspace{-14pt}
\includegraphics[width=0.875\columnwidth]%
 {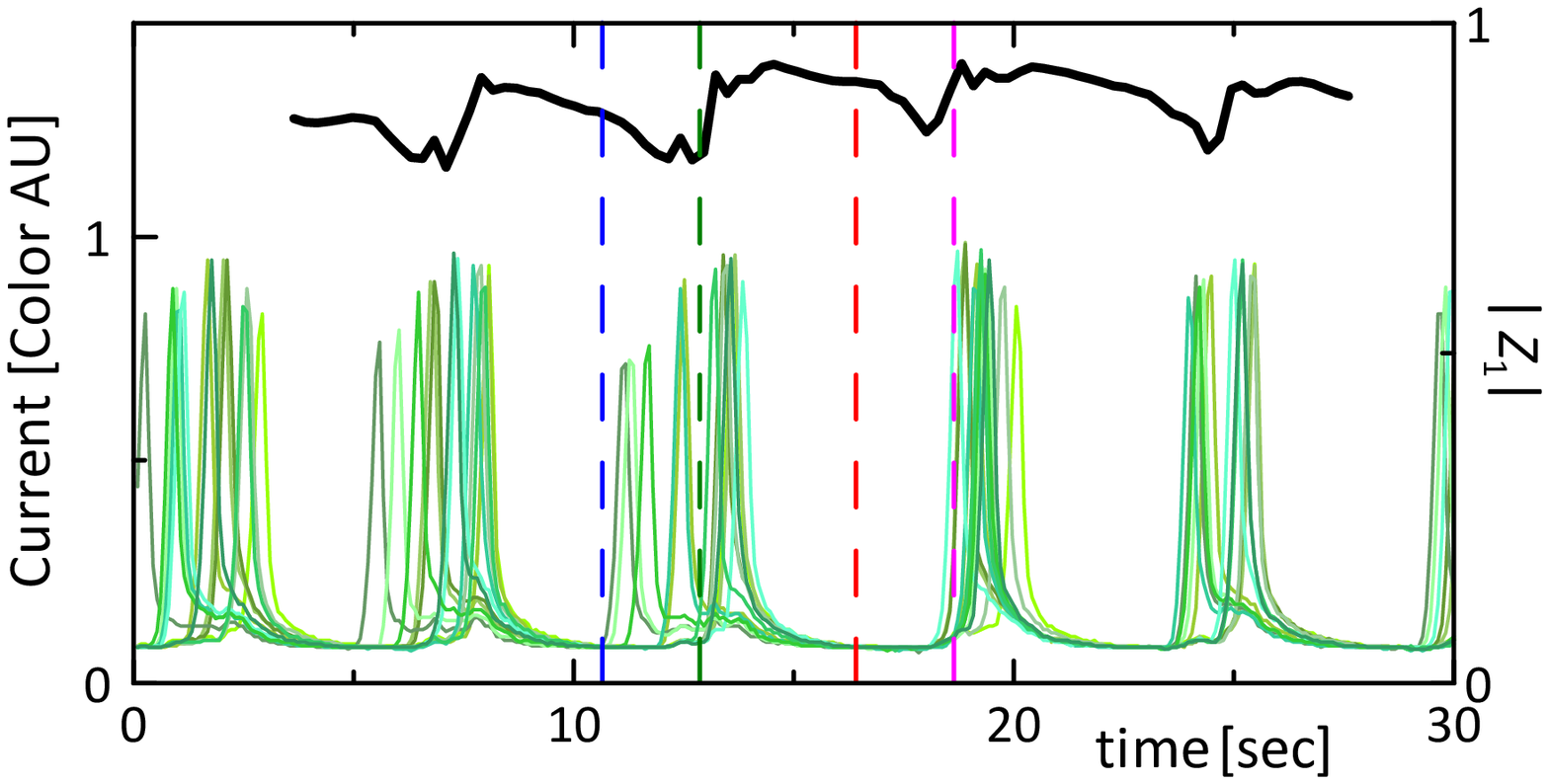}
\\[9pt]
(b)\hspace{-14pt}
\includegraphics[width=0.525\columnwidth]%
 {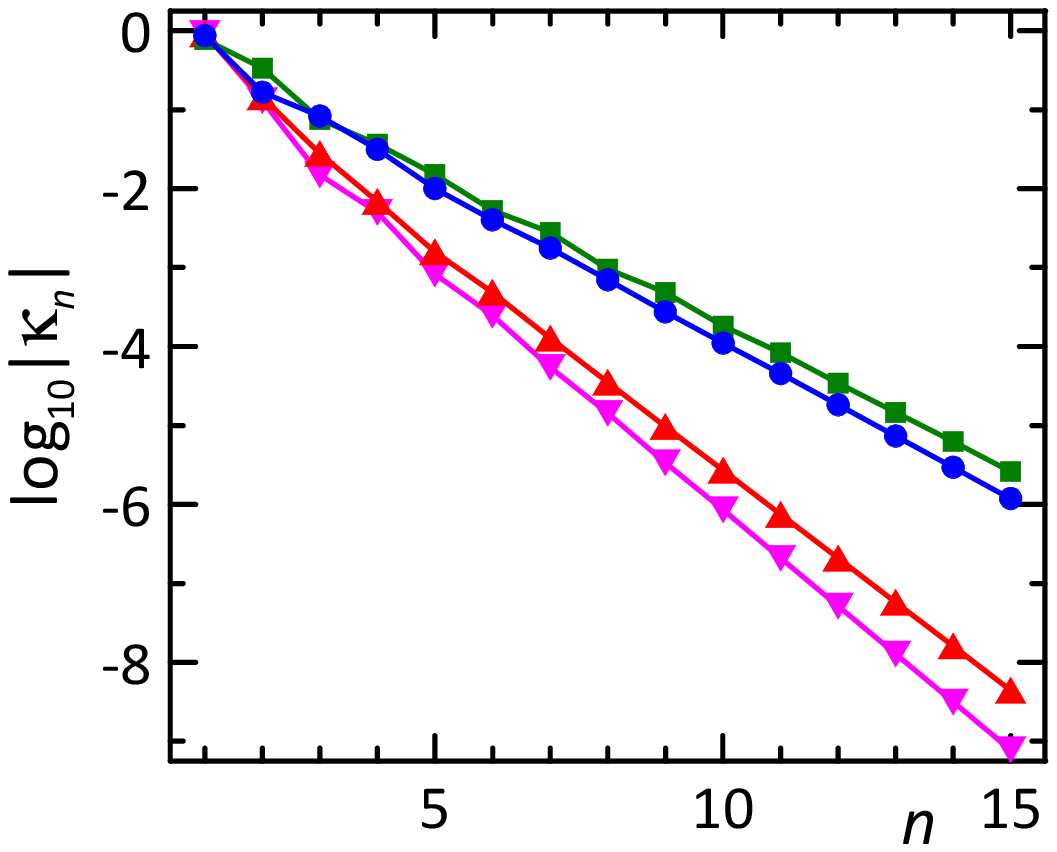}
}
\caption{Data processing for the population of $46$ electrochemical oscillators presented in Fig.~1c of Ref.~\cite{Kori-etal-2018}: The order parameter $|Z_1|$ is plotted with the bold black line in panel~(a); light-green lines show the reconstructed signals of each 5th oscillator. (b):~The first 15 circular cumulants are plotted for $t=10.7$ (blue circles), $12.9$ (green squares), $16.4$ (red up-pointed triangles), and $18.6$ (magenta down-pointed triangles); the time instants are marked with dashed lines in panel~(a).
}
  \label{fig5}
\end{figure}

As another example, let us consider experimental data for electrochemical oscillators~\cite{Kiss-etal-1999,Kori-etal-2018}.
For illustration, we intentionally analyze the data of different type as compared to the previous example. Instead of using raw measurements data, we digitize the shadowgraph in Fig.~1c of~\cite{Kori-etal-2018} with oscillation pattern of a population of $46$ coupled oscillators, where grayscale indicates the instantaneous current of an individual oscillator. The color is a one-to-one function of the current; therefore, such data are sufficient for calculation of the protophase (via Hilbert transform), the genuine phase, and the order parameters. In Fig.~\ref{fig5}, the reconstructed signals, the order-parameter $|Z_1(t)|$, and circular cumulant series are presented. The circular cumulant series clearly exhibit rapidly decaying geometric progressions.

%%%%%%%%%%%%%%%%%%%%%%%%%%%%%%%%%%%%%%%%%%%%%%%%%%%%%%%%%%%%
%%%%%%%%%%%%%%%%%%%%%%%%%%%%%%%%%%%%%%%%%%%%%%%%%%%%%%%%%%%%

\section{Conclusion}
For a variable on the circle we have derived that the circular cumulant series has either one nonzero element or an infinite number of them. The former corresponds to the wrapped Cauchy distribution {\it or\,} the Ott--Antonsen ansatz. With two or a larger but finite number of nonzero elements the high-order Kuramoto--Daido order parameters $\langle{e^{im\varphi}}\rangle$ tend to infinity, while their absolute value is not allowed to exceed $1$. This should be taken into account when one deals with the systems governed by non-trivial macroscopic variables like the firing rate in a network of quadratic integrate-and-fire neurons~\cite{Montbrio-Pazo-Roxin-2015}. Specifically, this firing rate [Eqs.~(\ref{eq:t304}), (\ref{eq:t303})] possesses good convergence properties on the Ott--Antonsen manifold and for all considered generic distributions on the circle, while it {\em always} diverges for any finite number of nonzero circular cumulants.

One can compare this situation with the case of a variable on the line, where the cumulant series has either only two first nonzero cumulants or an infinite number of them. The case of two nonzero cumulants corresponds to the Gaussian distribution. This apparent dissimilarity between linear and phase variables actually preserves a concordance between them: in both cases, the admitted truncation is characterised by two quantities. On the line, the first and second real-valued cumulants determine the centering and the width of a distribution, respectively, and on the circle, the argument and the absolute value of the first complex-valued circular cumulant determine the same characteristics of a distribution. On the line, the third and fourth cumulants quantify the distribution asymmetry and the deviation of tails from a reference law (kurtosis); on the circle, the argument and the absolute value of the second cumulant do the same.

For linear variables, in some cases, one is strictly bound to the Gaussian reduction; e.g., the Fokker--Planck equation corresponds to the white Gaussian noise and the analogues of this equation for a white noise accounting for a finite number of nonzero higher cumulants exhibit unphysical behavior. Meanwhile, in a wide range of problems, one can construct approximations accounting for corrections owned by a finite number of higher cumulants and benefit from them, or quantify the deviation from a Gaussian distribution with the third and forth cumulants (e.g.,~\cite{Peter-Pikovsky-2018}). Similarly, for phase variables, in some cases, one is allowed to deal with no finite circular cumulant truncation except the Ott--Antonsen ansatz; see the example of neuron firing rate. Simultaneously, in a range of problems, the approximations accounting for higher cumulant contributions yield accurate solutions where the Ott--Antonsen ansatz fails~\cite{Tyulkina-etal-2018,Goldobin-etal-2018}.

Strictly speaking, the example of neuron firing rate does not match the example of the Fokker--Planck equation and its analogues for a finite number of higher cumulants. With the latter, the issue cannot be handled, while the former case can be handled in a regular way for practical situations. For the case of a geometric progression, $|\kappa_n|\propto\varepsilon^{n-1}$, $\varepsilon\ll1$, one should not use the exact formulae for a finite number $N$ of circular cumulants, but use expansions with terms up to $\kappa_n^{(N-1)/(n-1)}$ and $n\le N$.

We have examined the relevance of the geometric progression allowing for a regular approach to constructing approximations of a prescribed accuracy or with a predefined number of circular cumulants. This progression is always present for wrapped Gaussian, von Mises, and non-Cauchy heavy-tail distributions~(\ref{eq:wnC}), and is typical in experiments, as demonstrated with data for coupled biological~\cite{Abel-etal-2016} and electrochemical oscillators~\cite{Kiss-etal-1999,Kori-etal-2018}.

\begin{acknowledgments}
The authors are thankful to Arkady Pikovsky, Lyudmila Klimenko, and Robert Banks for discussions and useful comments and acknowledge the financial support from a joint RSF-DFG project (Russian Science Foundation grant no.\ 19-42-04120).
\end{acknowledgments}

\appendix
\section{Derivation of two-cumulant truncation}
\label{asec:2cZm}
When only two first circular cumulants are nonzero, the circular cumulant generating function $\Psi(k)=\kappa_1k+\kappa_2k^2$, and --- since $\Psi(k)=k\partial_k\ln{F(k)}$, where the moment generating function $F(k)=1+Z_1k+Z_2\frac{k^2}{2!}+Z_3\frac{k^3}{3!}+\dots$ (see~\cite{Tyulkina-etal-2018}) ---  $\ln{F}=\kappa_1k+\kappa_2\frac{k^2}{2}$. Thus,
\begin{align}
&\textstyle
F=e^{\kappa_1k}e^{\kappa_2\frac{k^2}{2}}
%\nonumber\\
%&\quad\textstyle
%{}
=\sum\limits_{m_1=0}^{\infty}\kappa_1^{m_1}\frac{k^{m_1}}{m_1!} \sum\limits_{m_2=0}^{\infty}\big(\frac{\kappa_2}{2}\big)^{m_2}\frac{k^{2m_2}}{m_2!}\,.
\label{eq:a101}
\end{align}
Gathering the terms with $k^m$ in product~(\ref{eq:a101}) for $m=2n$, one finds
\[
\textstyle
\frac{Z_{2n}}{(2n)!}=\frac{\kappa_1^{2n}}{(2n)!}+\frac{\kappa_1^{2n-2}}{(2n-2)!}\frac{\kappa_2}{2\cdot1!} %+\frac{\kappa_2^{n-2}}{2^{n-2}(2n-4)!}\frac{\kappa_1^4}{4!}
+\dots+\frac{\kappa_2^n}{2^nn!}\,,
\]
and
\begin{align}
\textstyle
Z_{2n}=\sum_{j=0}^{n}\frac{(2n)!}{(2n-2j)!j!}\frac{\kappa_1^{2n-2j}\kappa_2^{j}}{2^j}\,.
\label{eq:a102}
\end{align}
For $m=2n+1$,
\[
\textstyle
\frac{Z_{2n+1}}{(2n+1)!}=\frac{\kappa_1^{2n+1}}{(2n+1)!}+\frac{\kappa_1^{2n-1}}{(2n-1)!}\frac{\kappa_2}{2\cdot1!} %+\frac{\kappa_2^{n-2}}{2^{n-2}(2n-4)!}\frac{\kappa_1^5}{5!}
+\dots+\frac{\kappa_1}{1!}\frac{\kappa_2^n}{2^nn!}\,,
\]
and
\begin{align}
\textstyle
Z_{2n+1}=\sum_{j=0}^{n}\frac{(2n+1)!}{(2n+1-2j)!j!}\frac{\kappa_1^{2n+1-2j}\kappa_2^j}{2^j}\,.
\label{eq:a103}
\end{align}

For $m\gg|\kappa_1|/\sqrt{|\kappa_2|}$, the second term in sums~(\ref{eq:a102}) and (\ref{eq:a103}) is large compared to the first one.
For $m\gg\sqrt{|\kappa_2|}/|\kappa_1|$, the term ahead of the last term in these sums is large compared to the last one.
Hence, for $m\gg \max(\sqrt{|\kappa_2|}/|\kappa_1|,|\kappa_1|/\sqrt{|\kappa_2|}) \sim\frac{\sqrt{|\kappa_2|}}{|\kappa_1|}+\frac{|\kappa_1|}{\sqrt{|\kappa_2|}}$, the leading contributions into the sum are owned by the terms which are far from the sum edges. For large $n$, $j$, and $(n-j)$, one can employ the Stirling's approximation, $n!\approx\sqrt{2\pi n}(n/e)^n$, and find
$\frac{m!}{(m-2j)!j!}\approx\sqrt{\frac{m}{2\pi(m-2j)j}}\frac{m^m}{(m-2j)^{m-2j}j^je^j}$ and
\begin{align}
&\textstyle
Z_m\approx\sum_{j=0}^{m/2}s_{m;j}\,,
\label{eq:a104a}
\\
&\textstyle
s_{m;j}\equiv
\sqrt{\frac{m}{2\pi(m-2j)j}} \frac{m^m \kappa_1^{m-2j}\kappa_2^j}{(m-2j)^{m-2j}(2ej)^j}.
\label{eq:a104b}
\end{align}

The largest contribution into $Z_m$ is made by the summand $s_{m;l}$, for which $\frac{\mathrm{d}}{\mathrm{d}j}\ln{|s_{m;j}|}=0$\,:
\begin{align}
\textstyle
\frac{\mathrm{d}}{\mathrm{d}l}\left(-\big(m-2l+\frac12\big)\ln(m-2l)
 -\big(l+\frac12\big)\ln(2l)\right.
\nonumber\\
\textstyle\left.{}-l+l\ln|\kappa_2|+(m-2l)\ln|\kappa_1|\right)
\nonumber\\
\textstyle
{}=\ln\left(\frac{(m-2l)^2}{2l}\frac{|\kappa_2|}{|\kappa_1|^2}\right)+\frac{1}{m-2l}-\frac{1}{2l}=0\,.
\label{eq:a105}
\end{align}

Recall, these evaluations are valid for $m\gg a^{-1/2}+a^{1/2}$, where we introduce notation
\[
a=\frac{|\kappa_1|^2}{2|\kappa_2|}\,.
\]
For simplicity of calculations, we conduct further consideration for even larger
\[
m\gg M_3=a^{-3}+a^3,
\]
which allows one to rigorously neglect as many contributions in expansions as possible.

For large $m$, one can solve Eq.~(\ref{eq:a105}) iteratively. At the first iteration, one can assume the logarithm argument to be 1 and find
\[
2l^{(1)}=m+a-\sqrt{a^2+2ma}=m-\sqrt{2ma}+a+o(1)\,.
\]
Substitution of $l=l^{(1)}$ into the second and third terms of Eq.~(\ref{eq:a105}) yields
\begin{align}
&\textstyle
(m-2l)^2=2a\Big(1+\frac{1}{2l^{(1)}}-\frac{1}{m-2l^{(1)}}
\nonumber\\
&\textstyle\qquad\qquad\qquad
{}+\frac12\left(\frac{1}{2l^{(1)}}-\frac{1}{m-2l^{(1)}}\right)^2+\dots\Big)2l
\nonumber\\
&\textstyle\qquad
{}=2\left(a-\sqrt{\frac{a}{2m}}+\frac{a}{2m}+\frac{1}{4m} +O\left(\frac{a^{3/2}}{m^{3/2}}\right)\right)2l\,,
\nonumber
\end{align}
and
\begin{equation}
2l=m-\sqrt{2ma}+a+\frac12+o(1)\,.
\label{eq:a106}
\end{equation}

Further,
\begin{align}
&\textstyle
|s_{m;l}|\approx \frac{1}{\sqrt{\pi}(2ma)^\frac14} \left(\frac{m}{m-2l}\right)^{m-2l} \left(\frac{m^2}{2el}\right)^l |\kappa_1|^{m-2l}|\kappa_2|^l
\nonumber\\
&\textstyle\qquad
=\frac{|\kappa_1|^m}{\sqrt{\pi}(2ma)^\frac14} \left(\frac{m}{\sqrt{2ma}-a-\frac12-o(1)}\right)^{m-2l}
\nonumber\\
&\textstyle\qquad\qquad
{}\times\left(\frac{m^2}{m-\sqrt{2ma}+a+\frac12+o(1)}\frac{|\kappa_2|}{|\kappa_1|^2}\right)^l
\nonumber\\
&\textstyle\qquad
=\frac{|\kappa_1|^mm^\frac{m}{2}}{\sqrt{\pi}(2ma)^\frac14(2a)^\frac{m-2l}{2}} \Big(1-\sqrt{\frac{a}{2m}}-\frac{\frac12+o(1)}{\sqrt{2ma}}\Big)^{-m+2l}
\nonumber\\
&\textstyle\qquad\qquad
{}\times
\Big(1-\sqrt{\frac{2a}{m}}+\frac{a}{m}+\frac{\frac12+o(1)}{m}\Big)^{-l} \left(\frac{1}{2ea}\right)^l.
\nonumber
\end{align}
For $(1+\varepsilon)^N$, one can employ
\begin{equation}
\ln(1+\varepsilon)^N=\varepsilon N-\frac{\varepsilon^2}{2}N+\frac{\varepsilon^3}{3}N-\dots
\label{eq:a107}
\end{equation}
and calculate
\begin{align}
&\textstyle
\Big(1-\sqrt{\frac{a}{2m}}-\frac{\frac12+o(1)}{\sqrt{2ma}}\Big)^{-\sqrt{2ma}+a+\frac12+o(1)}
\nonumber\\
&\textstyle\qquad
{}=e^{a+\frac12+o(1)-\sqrt{\frac{a^3}{2m}}+\dots+\sqrt{\frac{a^3}{8m}}+\dots}
=e^{a+\frac12+o(1)},
\nonumber
\end{align}
\begin{align}
&\textstyle
\Big(1-\sqrt{\frac{2a}{m}}+\frac{a}{m} +\frac{\frac12+o(1)}{m}\Big)^{-\frac{m-\sqrt{2ma}+a+\frac12+o(1)}{2}}
\nonumber\\
&\textstyle
{}=\exp\Big[\sqrt{\frac{ma}{2}}-a+\sqrt{\frac{a^3}{2m}}+\dots -\frac{a}{2}+\sqrt{\frac{a^3}{2m}}+\dots
\nonumber\\
&\textstyle
{}-\frac14-o(1)+\dots +\frac12\Big(\frac{2a}{m}-\big(\frac{2a}{m}\big)^\frac32 +\dots\Big)\frac{m-\sqrt{2ma}+\dots}{2}
\nonumber\\
&\textstyle\qquad
{}+\frac13\sqrt{\frac{2a^3}{m}}+\dots\Big]
{}=e^{\sqrt{\frac{ma}{2}}-a-\frac14+o(1)}.
\nonumber
\end{align}
Hence,
\begin{align}
&\textstyle
|s_{m;l}|=\frac{|\kappa_1|^mm^\frac{m}{2}}{\sqrt{\pi}(2ma)^\frac14(2a)^\frac{m}{2} e^{\frac{m}{2}-\sqrt{2ma}+\frac{a}{2}}}[1+o(1)]
\nonumber\\
&\textstyle\qquad
{}=\frac{1}{\sqrt{\pi}(2ma)^\frac14} \left(\frac{m|\kappa_2|}{e}\right)^{\frac{m}{2}}e^{\sqrt{2ma}-\frac{a}{2}}[1+o(1)]
%\nonumber\\
%&\textstyle
%{}=\sqrt{\frac{|\kappa_2|}{\pi|\kappa_1|}} %\left(\frac{m|\kappa_2|}{e}\right)^{\frac{m}{2}-\frac14}e^{\frac{\sqrt{m}|\kappa_1|}{\sqrt{|\kappa_2|}} -\frac{|\kappa_1|^2+|\kappa_2|}{4|\kappa_2|}}[1+o(1)].
\,.
\label{eq:a108}
\end{align}

Let us now consider the vicinity of $j=l$. From Eq.~(\ref{eq:a104b}),
\begin{align}
&\textstyle
\frac{s_{m;l+r}}{s_{m;l}}
=\sqrt{\frac{(m-2l)l}{(m-2l-2r)(l+r)}}
\frac{\kappa_1^{-2r}\kappa_2^r(m-2l)^{m-2l}(2el)^l}{(m-2l-2r)^{m-2l-2r}(2e(l+r))^{l+r}}
\nonumber\\
&\textstyle\qquad
=\left(1-\frac{2r}{m-2l}\right)^{-\frac12}\left(1+\frac{2r}{2l}\right)^{-\frac12} \left(\frac{\kappa_2}{\kappa_1^2}\right)^r
\nonumber\\
&\textstyle\qquad\qquad
\times(m-2l)^{2r}\left(1-\frac{2r}{m-2l}\right)^{-m+2l+2r}
\nonumber\\
&\textstyle\qquad\qquad
\times(2el)^{-r}\left(1+\frac{r}{l}\right)^{-l-r}\,.
\nonumber
\end{align}
Employing expansion (\ref{eq:a107}), one can find
\begin{align}
&\textstyle
\frac{s_{m;l+r}}{s_{m;l}}\approx\left(\frac{\kappa_2}{\kappa_1^2}\frac{(m-2l)^2}{2el}\right)^r
\exp\Big[\frac{2r(m-2l-2r)}{m-2l}
\nonumber\\
&\textstyle\quad
{}+\frac12\frac{4r^2(m-2l-2r)}{(m-2l)^2}+\dots -\frac{r(l+r)}{l}+\frac12\frac{r^2(l+r)}{l^2}+\dots\Big]
\nonumber\\
&\textstyle\qquad\qquad
\approx\left(\frac{\kappa_2}{\kappa_1^2}\frac{(m-2l)^2}{2l}\right)^r
e^{-\frac{2r^2}{m-2l}-\frac{r^2}{2l}}
\nonumber\\
&\textstyle\qquad\qquad
\approx e^{i\Theta r-\sqrt{\frac{2}{ma}}r^2},
\label{eq:a109}
\end{align}
where we used that the logarithm argument in Eq.~(\ref{eq:a105}) is close to 1, and introduced notation
\[
e^{i\Theta}\equiv\frac{\kappa_2}{|\kappa_2|}\frac{|\kappa_1|^2}{\kappa_1^2}\,.
\]

%%%%%%%%%%%%%%%%%%%%%%%%%%%%%%%%%%%%%%%%%%%%%%%%%%%%%%%%%%%%
\begin{figure}[!t]
\center{\footnotesize\sf
(a)\hspace{-13pt}
\includegraphics[width=0.477\columnwidth]%
 {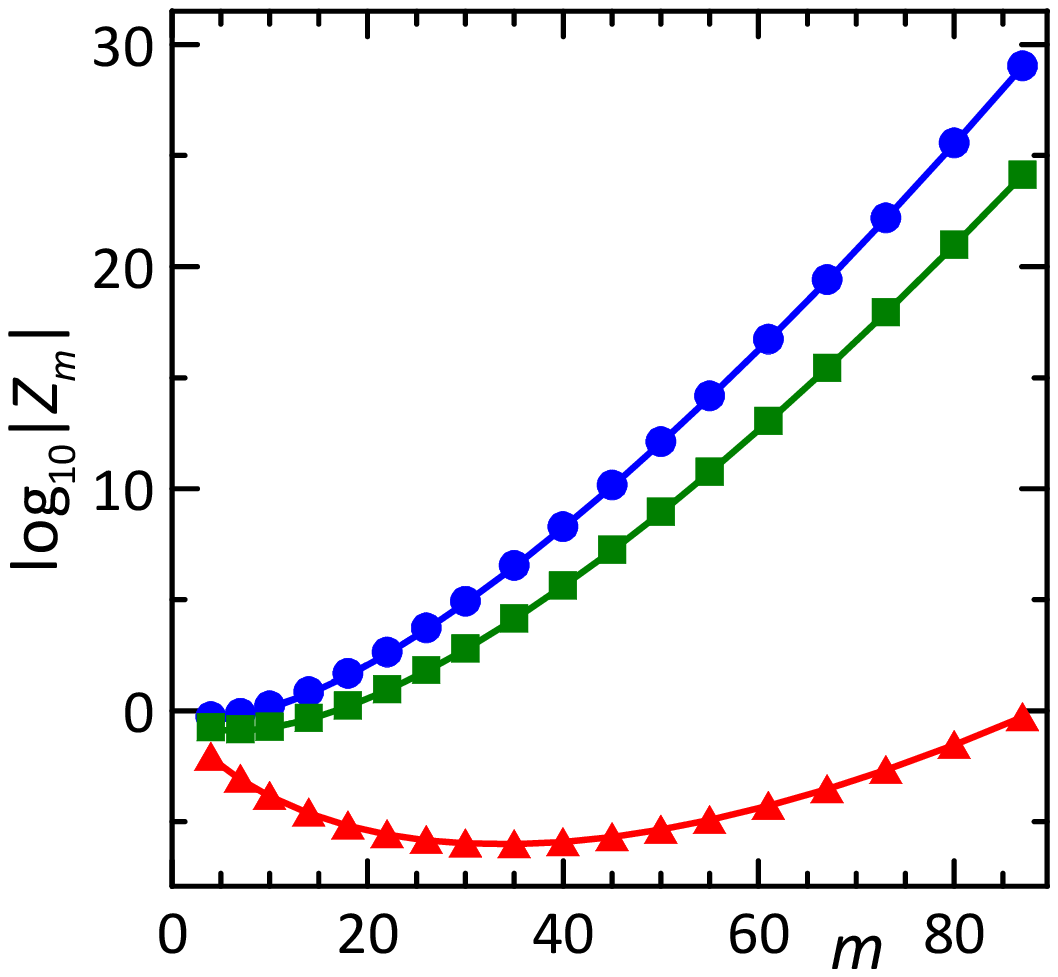}
\;
(b)\hspace{-13pt}
\includegraphics[width=0.477\columnwidth]%
 {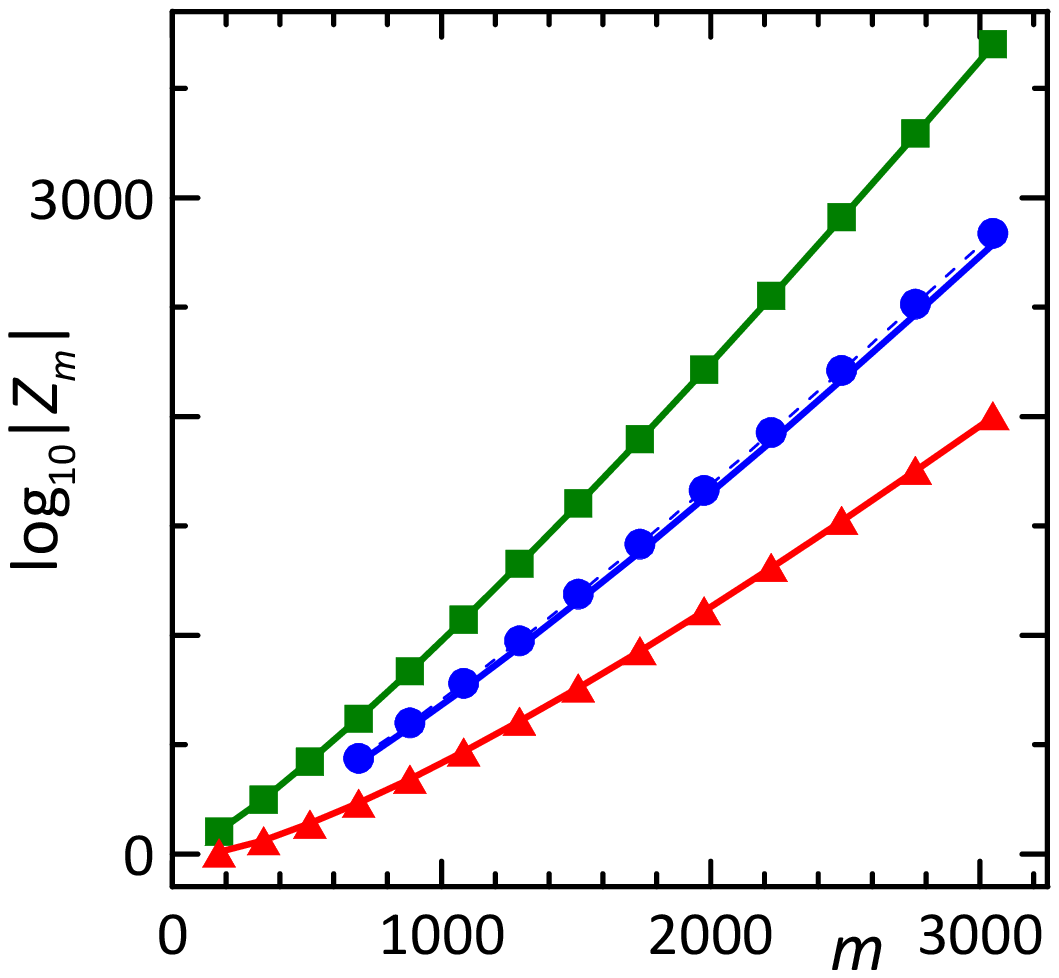}
\\[9pt]
(c)\hspace{-13pt}
\includegraphics[width=0.477\columnwidth]%
 {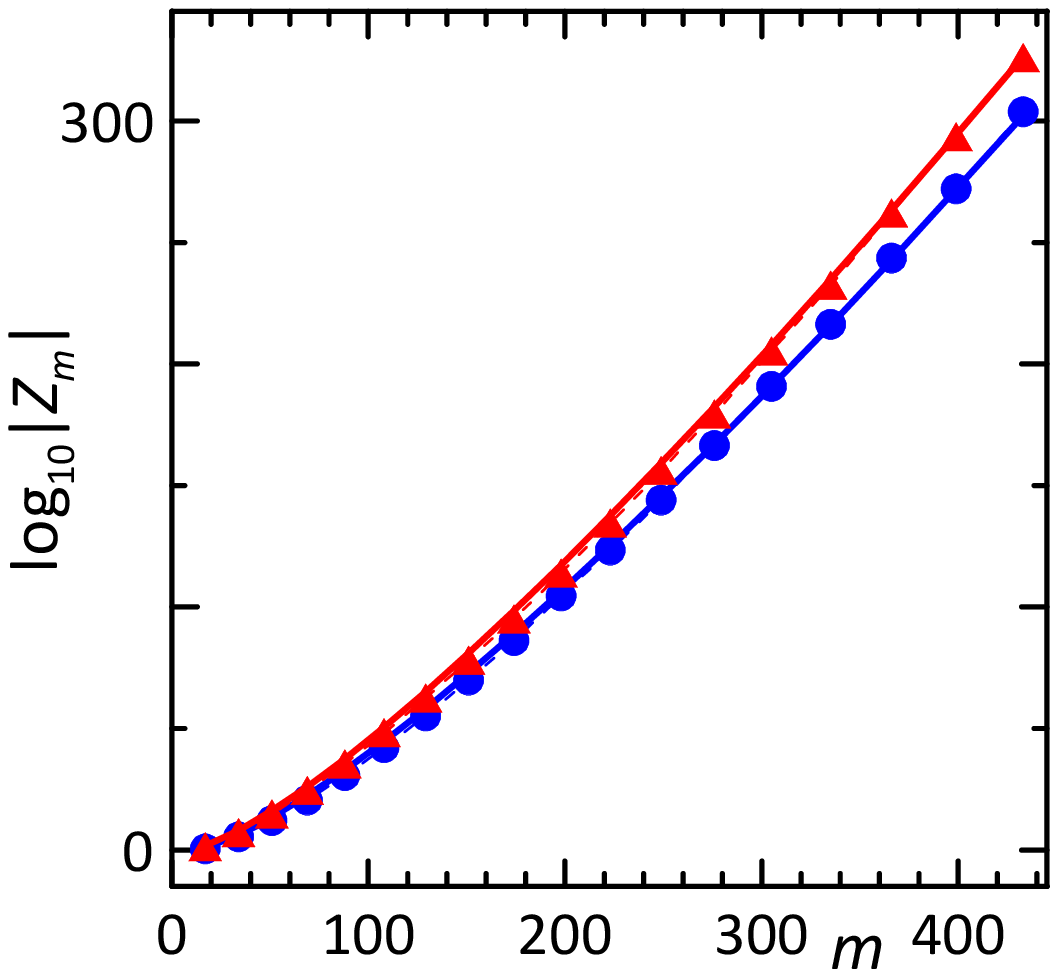}
\;
(d)\hspace{-13pt}
\includegraphics[width=0.477\columnwidth]%
 {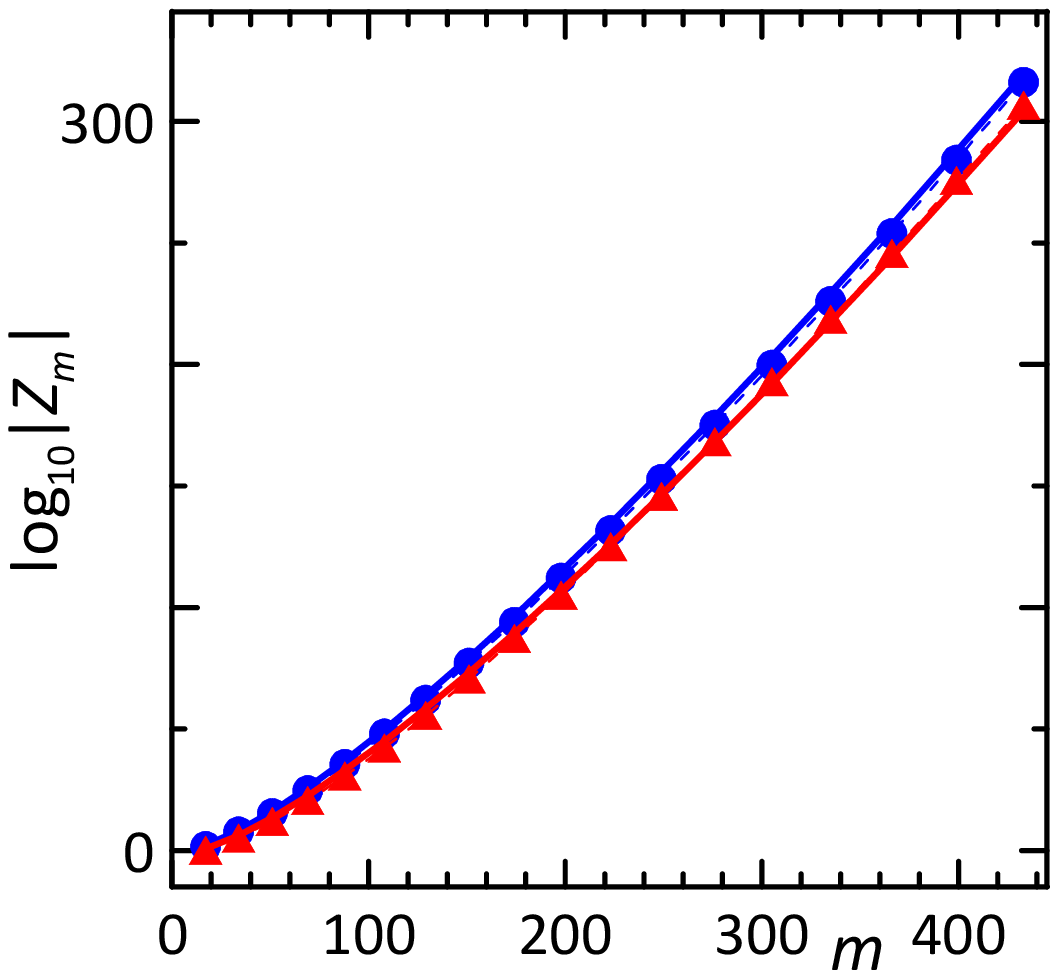}
}
\caption{Asymptotic formula~(\ref{eq:t110}) and the exact sum~(\ref{eq:t1023}) are plotted with lines and symbols, respectively, for a two-cumulant truncation for $\Theta=0$ in panel~(a) and finite $\Theta\ne0$ in panels~(b-d).
(a):~For the wrapped heavy-tail non-Cauchy distribution~(\ref{eq:wnC}) with $\mu=-0.25$ (left panel in Fig.~\ref{fig3}), $\Theta=0$; $\sigma=0.1$ (blue circles), $0.32$ (green squares), $1$ (red triangles).
(b):~For the wrapped Gaussian distribution~(\ref{eq:wG}), the even-order circular cumulants are negative, which means $\Theta=\pi$; $\sigma=0.25$ (blue circles), $1$ (green squares), $2$ (red triangles).
(c):~SCN2 data set for biological oscillators (Fig.~\ref{fig4} b,d) at $t=280$ (blue circles) and $368$ (red triangles), where $\Theta\approx-0.93\pi$.
(d):~Electrochemical oscillators (Fig.~\ref{fig5}) at $t=10.7$ (blue circles) and $16.4$ (red triangles), where $\Theta\approx-0.75\pi$ and $-0.97\pi$, respectively.
}
  \label{fig6}
\end{figure}
%%%%%%%%%%%%%%%%%%%%%%%%%%%%%%%%%%%%%%%%%%%%%%%%%%%%%%%%%%%%

With Eq.~(\ref{eq:a109}), one can assess the order of magnitude of the sum in $Z_m$ as an integral;
\[
\textstyle
Z_m\approx s_{m;l}\int\limits_{-\infty}^{+\infty}\frac{s_{m;l+r}}{s_{m;l}}\mathrm{d}r
=\sqrt{\pi}\left(\frac{ma}{2}\right)^\frac14 e^{-\frac{\sqrt{2ma}\Theta^2}{16}}s_{m;l}\,.
\]
Substituting $|s_{m;l}|$ from Eq.~(\ref{eq:a108}), one finally finds
\begin{align}
&\textstyle
|Z_m|\approx
\frac{1}{\sqrt{2}}\left(\frac{m|\kappa_2|}{e}\right)^\frac{m}{2}
e^{\sqrt{2ma}\left(1-\frac{\Theta^2}{16}\right)-\frac{a}{2}}
\nonumber\\
&\textstyle\qquad
=\frac{1}{\sqrt{2}}\left(\frac{m|\kappa_2|}{e}\right)^\frac{m}{2}
e^{\frac{\sqrt{m}|\kappa_1|}{\sqrt{|\kappa_2|}}\left(1-\frac{\Theta^2}{16}\right) -\frac{|\kappa_1|^2}{4|\kappa_2|}}.
\label{eq:a110}
\end{align}
The transition from a sum to an integral introduces inaccuracy. However, one can see in Fig.~\ref{fig6} that, for large $m$, the results with an integral match the exact sum satisfactorily not only for $\Theta=0$ (where the accuracy is high already for non-large $m$), but also for $\Theta$ as large as $\sim\pi$.

From Eq.~(\ref{eq:a110}), which is valid for $m\gg M_3$, one can see that for $m>e/|\kappa_2|$, the absolute value $|Z_m|$ becomes larger than $1$. However, this is not possible for the average value $\langle{e^{im\varphi}}\rangle$ of a phase $\varphi$ on the circle.

\section{Derivation of truncation of $\kappa_{m}$ for $m>N$}
\label{asec:NcZm}
For nonzero $\kappa_m$, $m=1,2,...,N$,
\begin{align}
%\textstyle
F(k)=\exp\left(\sum_{m=1}^N\kappa_m\frac{k^m}{m}\right)
=\prod\limits_{m=1}^Ne^\frac{\kappa_mk^m}{m}
\nonumber\\
%\textstyle
=\prod\limits_{m=1}^N\sum_{j_m=0}^\infty\left(\frac{\kappa_m}{m}\right)^{j_m}\frac{k^{m j_m}}{(j_m)!}\,.
\nonumber
\end{align}
Hence,
\begin{equation}
%\textstyle
Z_m=\!\!\!\sum\limits_{j_1+2j_2+\dots \atop {}+Nj_N=m}\!
 \frac{m!}{j_1!j_2!\dots j_N!}\left(\frac{\kappa_1}{1}\right)^{j_1} \left(\frac{\kappa_2}{2}\right)^{j_2}\!\!\dots
 \left(\frac{\kappa_N}{N}\right)^{j_N}\!\!.
\label{eq:a201}
\end{equation}
Again, for sufficiently large $m$, the principal contributions are owned by the summands which are far from the boundaries of the summation domain in the index space. In this section we assume $m$ to be large compared to any reference value for it. For large $m$, $j_1$, $j_2$,..., $j_N$, one can use the Stirling's formula and replace the summation with integration over the hyperplane $j_1+2j_2+3j_3+\dots+Nj_N=m$. Thus,
\begin{align}
&\textstyle
Z_m\approx\int\mathrm{d}j_2\int\mathrm{d}j_3\dots\int\mathrm{d}j_{N} s_{m;j_1j_2\dots j_N}\,,
\label{eq:a202a}
\\
&\textstyle
s_{m;j_1j_2\dots j_N}\equiv
\sqrt{\frac{m}{(2\pi)^{N-1}j_1j_2\dots j_N}}
\nonumber\\
&\textstyle\qquad\qquad\qquad
 \times\frac{m^m\kappa_1^{j_1}\kappa_2^{j_2}\dots\kappa_N^{j_N}}{(j_1)^{j_1}(2ej_2)^{j_2} \dots(Ne^{N-1}j_N)^{j_N}},
\label{eq:a202b}
\\
&\textstyle\qquad
j_1=m-2j_2-3j_3-\dots-Nj_N\,.
\nonumber
\end{align}

One can find the maximum of $|s_{m;j_1j_2\dots j_N}|$ on the hyperplane by means of the method of Lagrange multipliers. Equations
\begin{align}
&\textstyle
\frac{\partial}{\partial l_n}\Big(\ln|s_{m;l_1l_2\dots l_N}|+\lambda\!\!\sum\limits_{n^\prime=1}^Nn^\prime l_{n^\prime}\Big)=0\,,
%\nonumber\\
%\textstyle
\quad
n=1,\dots,N\,,
\nonumber\\
&\textstyle\qquad
\sum\limits_{n=1}^Nn l_n=m
\nonumber
\end{align}
take the form
\begin{align}
&\textstyle
\frac{1}{2l_n}-\ln\frac{|\kappa_n|}{nl_n}=n(\lambda-1)\,,
%\nonumber\\
%\textstyle
\quad
n=1,\dots,N\,,
\nonumber\\
&\textstyle\qquad
\sum\limits_{n=1}^Nn l_n=m\,.
\nonumber
\end{align}

To the leading order, one finds
\begin{equation}
\textstyle
%nl_n\approx\frac{|\kappa_n|m^\frac{n}{N}}{|\kappa_N|^\frac{n}{N}} \left(1-\frac{n(N-1)}{4m}\right) \left(1-\frac{n}{Nm^\frac{1}{N}}\frac{|\kappa_{N-1}|}{|\kappa_N|^{\frac{N-1}{N}}}\right)+\frac{n}{2}\,,
nl_n\approx|\kappa_n|\Lambda^n-\frac{n}{2}\,,
\label{eq:a203}
\end{equation}
where $\Lambda\equiv e^{\lambda-1}$ obeys
\[
\textstyle
|\kappa_N|\Lambda^N+|\kappa_{N-1}|\Lambda^{N-1}+\dots+|\kappa_{1}|\Lambda=m+\frac{N(N-1)}{4}\,;
\]
therefore,
\begin{equation}
\textstyle
\Lambda\approx\left(\frac{m}{|\kappa_N|}\right)^\frac{1}{N} \left(1+\frac{N-1}{4m}-\frac{1}{Nm^\frac{1}{N}}\frac{|\kappa_{N-1}|}{|\kappa_N|^{1-\frac{1}{N}}}\right).
\label{eq:a204}
\end{equation}
Hence, Eq.~(\ref{eq:a203}) yields
\begin{equation}
\textstyle
nl_n\approx\frac{|\kappa_n|}{|\kappa_N|^\frac{n}{N}}m^\frac{n}{N} -\frac{n}{N}\frac{|\kappa_n||\kappa_{N-1}|}{|\kappa_N|^\frac{N+n-1}{N}}m^\frac{n-1}{N}\,,
\label{eq:a205}
\end{equation}
and
\begin{align}
&|s_{m;l_1l_2\dots l_N}|
\approx\sqrt{\frac{m}{(2\pi)^{N-1}l_1l_2\dots l_N}}
\nonumber\\
&\qquad\qquad
\times\prod_{n=1}^{N}
\left(\frac{1}{e^{n-1}\Lambda^n}\right)^{l_n}
\left(1+\frac{n}{2|\kappa_n|\Lambda^n}\right)^{l_n}
\nonumber\\
&\qquad
\approx\sqrt{\frac{m}{(2\pi)^{N-1}l_1l_2\dots l_N}}
\frac{e^{\sum_{n=1}^{N}l_n}}{e^m\Lambda^m}e^\frac{N}{2}
\nonumber\\
&\qquad
\approx\sqrt{\frac{m}{(2\pi)^{N-1}l_1l_2\dots l_N}}
\left(\frac{m^{1-\frac{1}{N}}|\kappa_N|^\frac{1}{N}}{e^{1-\frac{1}{N}}}\right)^m
\nonumber\\
&\qquad\qquad
\times\exp\left[\frac{m^{1-\frac{1}{N}}|\kappa_{N-1}|}{(N-1)|\kappa_N|^{1-\frac{1}{N}}}+O(m^{1-\frac{2}{N}})\right].
\label{eq:a206}
\end{align}
Further,
\begin{align}
\frac{s_{m;l_1+r_1\dots l_N+r_N}}{s_{m;l_1\dots l_N}}\approx\prod_{n=1}^N\kappa_n^{r_n} \frac{(ne^{n-1}l_n)^{l_n}}{\left(ne^{n-1}(l_n+r_n)\right)^{l_n+r_n}}
\nonumber\\
\approx\prod_{n=1}^N\left(\frac{\kappa_n}{e^{n-1}nl_n}\right)^{r_n} \left(1+\frac{r_n}{l_n}\right)^{-(l_n+r_n)},
\nonumber
\end{align}
where $r_1$ is dictated by $r_2$, ..., $r_N$ by virtue of the condition $\sum_{n=1}^{N}nr_n=0$. Employing expansion~(\ref{eq:a107}) and condition $\sum_{n=1}^{N}nr_n=0$, one can obtain
\begin{align}
&\frac{s_{m;l_1+r_1\dots l_N+r_N}}{s_{m;l_1\dots l_N}}
\approx\prod_{n=1}^N\left(\frac{\kappa_n}{e^{n-1}nl_n}\right)^{r_n}
\nonumber\\
&\times e^{-\frac{r_n}{l_n}(l_n+r_n)+\frac{r_n^2}{2l_n^2}(l_n+r_n)+\dots}
%\nonumber\\
=\prod_{n=1}^N\left(\frac{\kappa_n}{nl_n}\right)^{r_n}
 e^{-\frac{r_n^2}{2l_n}+\dots}
\nonumber\\
&\qquad\qquad
=\prod_{n=1}^N\left(\frac{\kappa_n}{|\kappa_n|}\right)^{r_n}\Lambda^{-nr_n}
 e^{-\frac{r_n^2}{2l_n}+\dots}
\nonumber\\
&\qquad\qquad\qquad
=\prod_{n=1}^N
 e^{i\alpha_nr_n-\frac{r_n^2}{2l_n}+\dots}\,,
\label{eq:a207}
\end{align}
where $e^{i\alpha_n}\equiv\kappa_n/|\kappa_n|$.

In integral (\ref{eq:a202a}) over the hyperplane $\sum_{n=1}^Nnj_n=m$ one can replace $\int\mathrm{d}j_N$ with $N^{-1}\int\mathrm{d}j_1$.
Substituting $r_N=-\sum_{n=1}^{N-1}\frac{n}{N}r_n$ into Eq.~(\ref{eq:a207}) and employing Eq.~(\ref{eq:a205}) to find $1/l_{n<N}\gg1/l_{N}$, one can obtain
\begin{align}
\frac{s_{m;l_1+r_1\dots l_N+r_N}}{s_{m;l_1\dots l_N}}
\approx\prod_{n=1}^{N-1}
 e^{i\Theta_nr_n-\frac{r_n^2}{2l_n}+\dots}\,,
\label{eq:a208}
\end{align}
where $\Theta_n=\alpha_n-(n/N)\alpha_N$. Recasting Eq.~(\ref{eq:a202a}) as
\[
Z_m\approx \frac{s_{m;l_1l_2\dots l_N}}{N}\int\limits_{-\infty}^{+\infty}\!\mathrm{d}r_{N-1}\dots \int\limits_{-\infty}^{+\infty}\!\mathrm{d}r_1
 \frac{s_{m;l_1+r_1\dots l_N+r_N}}{s_{m;l_1\dots l_N}}\,,
\]
and evaluating integrals, one finds
\[
Z_m\approx\frac{s_{m;l_1l_2\dots l_N}}{N}\prod_{n=1}^{N-1}\sqrt{2\pi l_n}e^{-\frac{l_n\Theta_n^2}{4}}.
\]

Substituting Eq.~(\ref{eq:a206}), one finds
\begin{align}
&
|Z_m|\approx\frac{1}{\sqrt{N}}\left(\frac{m^{1-\frac{1}{N}}|\kappa_N|^\frac{1}{N}}{e^{1-\frac{1}{N}}}\right)^m
\exp\Bigg[\frac{m^{1-\frac{1}{N}}|\kappa_{N-1}|}{(N-1)|\kappa_N|^{1-\frac{1}{N}}}
\nonumber\\
&\qquad\qquad
\times \left(1-\frac{\Theta_{N-1}^2}{4}\right)
%\nonumber\\
%&\qquad\qquad\qquad
+O(m^{1-\frac{2}{N}})\Bigg].
\label{eq:a209}
\end{align}
Eq.~(\ref{eq:a209}) for $N=2$ is in agreement with Eq.~(\ref{eq:a110}) (by definition, $\Theta_2=\Theta/2$).

According to Eq.~(\ref{eq:a209}), for sufficiently large $m$ (specifically, $m\gg\frac{|\kappa_{N-1}|^N}{|\kappa_N|^{N-1}}+\frac{|\kappa_N|^{N-1}}{|\kappa_{N-1}|^N}$) and $m>e/|\kappa_N|^\frac{1}{N-1}$, one finds $|Z_m|>1$, which is not admitted for $\langle{e^{im\varphi}}\rangle$ of a phase variable $\varphi$.

\section{Calculation of firing rate for generic distributions}
\label{asec:r}
The calculation of the firing rate for several generic distributions we provide in Sec.~\ref{sec:hrch} is based on the representation of distribution $w(\varphi)$ by Fourier amplitudes $Z_m$, with emphasis on the convergence issue for series of $Z_m$. In this section, we verify the results of calculations in Sec.~\ref{sec:hrch} with an alternative approach, not involving $Z_m$.

For QIFs, the probability density of voltage variable $V$
\[
\widetilde{w}(V)=w(\varphi)\left|\frac{\mathrm{d}\varphi}{\mathrm{d}V}\right|=w(2\arctan{V})\frac{2}{1+V^2}
\]
[cf Eq.~(\ref{eq:V-phi})], and the firing rate (e.g., see~\cite{Montbrio-Pazo-Roxin-2015})
\begin{equation}
r(t)=\lim_{V\to\infty}\widetilde{w}(V)V^2=2w(\pi)\,.
\label{eq:a301}
\end{equation}

For wrapped Gaussian distribution~(\ref{eq:wG}),
the firing rate reads
\begin{equation}
r_\mathrm{wG}=\sum_{n=-\infty}^{+\infty}\frac{\sqrt{2}}{\sqrt{\pi}\sigma} e^{-\frac{(2\pi n-\psi-\pi)^2}{2\sigma^2}};
\label{eq:a302}
\end{equation}
for von Mises distribution~(\ref{eq:vM}),
\begin{equation}
r_\mathrm{vM}=\frac{e^{-\frac{\cos\psi}{\sigma^2}}}{\pi I_0(\sigma^{-2})};
\label{eq:a303}
\end{equation}
for wrapped non-Cauchy distribution~(\ref{eq:wnC}),
\begin{equation}
r_\mathrm{wnC}=\sum_{n=-\infty}^{+\infty}\frac{2\,\Gamma(1+\mu)}{\sqrt{\pi}\sigma\Gamma(\frac12+\mu) \left(1+\frac{\pi^2(2n+1)^2}{\sigma^2}\right)^{1+\mu}}.
\label{eq:a304}
\end{equation}

Eq.~(\ref{eq:a303}) for von Mises distribution, with the Jacobi--Anger expansion of $e^{a\cos(\alpha)}$, becomes identical to the real part of Eq.~(\ref{eq:WvM}):
 $\pi r_\mathrm{vM}=\mathrm{Re}\,W_\mathrm{vM}$.
Notice, for the imaginary part of Eq.~(\ref{eq:WvM}), such a simple representation of the sum in terms of elementary functions is not possible.

For wrapped Gaussian distribution, the sum~(\ref{eq:a302}) and the real part of Eq.~(\ref{eq:WwG}) are different sums; the former is a summation over branches of a wrapped function, while in the latter the summation is over Fourier modes. Both formulae involve the Gaussian function, because the Fourier transform of a Gaussian function is a Gaussian function. The multipliers ahead of $n^2/2$ in these sums are different, $(2\pi/\sigma)^2$ and $\sigma^2$, resulting in opposite dependencies of the convergence rate on the distribution width $\sigma$. This is natural: for sharp distributions with $\sigma\ll 1$, the wrapped branches introduce small corrections, but the Fourier series converges slow; for wide distributions, the contributions by wrapped branches decay slowly, but the Fourier series converges fast.
An analytical demonstration of the equivalence between Eqs.~(\ref{eq:a302}) and (\ref{eq:WwG}) [as well as between (\ref{eq:a304}) and (\ref{eq:WwnC})] is technically the derivation of $Z_m$ as provided in Sec.~\ref{sec:hrch}. However, the numerical calculation of (\ref{eq:a302}), (\ref{eq:a303}) and (\ref{eq:WwG}), (\ref{eq:WwnC}) provides us with a reliable test of the correctness of the presented formulae.

For wrapped non-Cauchy distribution~(\ref{eq:wnC}), the convergence of sum~(\ref{eq:WwnC}) corresponding to the Fourier representation is exponential, while the convergence of sum~(\ref{eq:a304}) corresponding to the physical-space representation is very slow; as $\mu$ tends to $-1/2+0$, the convergence of sum (\ref{eq:a304}) becomes infinitely slow.

\end{document}